\documentstyle[prl,aps,tighten,epsf,epsfig,pifont]{revtex}
\ifx\BORDER\relax\endinput\else\let\BORDER=\relax\fi 
\edef\borderelm{\the\catcode`\_} 
\catcode`\_=11

\def\hborder#1#2#3#4
   {\hbox to#1
	{\setbox0=\hbox{\borderelm#3}%
	 \borderelm#2%
	 \hskip0pt minus1fil %
	 \copy0\kern-.5\wd0 %
	 \cleaders\copy0\hskip0pt plus1fil %
	 \kern-.5\wd0\copy0 %
	 \hskip0pt minus1fil %
	 \borderelm#4%
	}%
   }

\def\border#1#2
   {\vbox to#1
        {\offinterlineskip\boxmaxdepth=\maxdimen
	 \dimen2=#2
	 \setbox0=\hbox to\dimen2{\borderelm3\hss\borderelm4}
	 \dimen0=\ht0 \advance\dimen0 by\dp0
	 \hborder{\dimen2}012
	 \vskip0pt minus1fil
	 \copy0\kern-.5\dimen0
	 \cleaders\copy0\vskip0pt plus1fil
	 \kern-.5\dimen0\copy0
	 \vskip0pt minus1fil
	 \hborder{\dimen2}567
	 \kern0sp
	}%
   }

\def\hxborder#1#2#3#4
   {\hbox to#1
	{\setbox0=\hbox{\borderelm#3}%
	 \borderelm#2%
	 \xleaders\copy0\hskip0pt plus1fil minus1fil %
	 \borderelm#4%
	}%
   }

\def\xborder#1#2
   {\vbox to#1
        {\offinterlineskip\boxmaxdepth=\maxdimen
	 \dimen2=#2
	 \setbox0=\hbox to\dimen2{\borderelm3\hss\borderelm4}
	 \hxborder{\diemn2}012
	 \xleaders\copy0\vskip0pt plus1fil minus1fil
	 \hxborder{\dimen2}567
	 \kern0sp
	}%
   }

\def\hminusborder#1#2#3#4
   {\hbox
	{\dimen0=#1\count2=\dimen0 %
	 \setbox0=\hbox{\borderelm#2}\advance\count2 by-\wd0 %
	 \setbox0=\hbox{\borderelm#4}\advance\count2 by-\wd0 %
	 \setbox0=\hbox{\borderelm#3}%
	 \advance\count2 by100 %
	 \divide\count2 by\wd0 %
	 \borderelm#2%
	 \cleaders\copy0\hskip\count2\wd0 %
	 \borderelm#4%
	}%
   }

\def\minus_border#1#2#3
   {\vbox
        {\offinterlineskip\boxmaxdepth=\maxdimen
	 \dimen2=#2\relax
	 \dimen0=#1\count2=\dimen0
	 \setbox0=\hminusborder{\dimen2}012
	 \advance\count2 by-\ht0 \advance\count2 by-\dp0
	 \setbox2=\hminusborder{\dimen2}567
	 \advance\count2 by-\ht2 \advance\count2 by-\dp2
	 \if.#3.\setbox4=\hbox to\wd0{\borderelm3\hss\borderelm4}
	 \else  \setbox4=\hminusborder{\dimen2}384
	 \fi
	 \advance\count2 by100
	 \dimen0=\ht4\advance\dimen0 by\dp4
	 \divide\count2 by\dimen0
	 \box0
	 \cleaders\copy4\vskip\count2\dimen0
	 \box2
	 \kern0sp
	}%
   }

\def\hplusborder#1#2#3#4
   {\hbox
	{\dimen0=#1\count2=\dimen0 %
	 \setbox0=\hbox{\borderelm#2}\advance\count2 by-\wd0 %
	 \setbox0=\hbox{\borderelm#4}\advance\count2 by-\wd0 %
	 \setbox0=\hbox{\borderelm#3}%
	 \advance\count2 by\wd0\advance\count2 by-100 %
	 \divide\count2 by\wd0 %
	 \borderelm#2%
	 \cleaders\copy0\hskip\count2\wd0 %
	 \borderelm#4%
	}%
   }

\def\plus_border#1#2#3
   {\vbox
        {\offinterlineskip\boxmaxdepth=\maxdimen
	 \dimen2=#2\relax
	 \dimen0=#1\relax\count2=\dimen0
	 \setbox0=\hplusborder{\dimen2}012
	 \advance\count2 by-\ht0 \advance\count2 by-\dp0
	 \setbox2=\hplusborder{\dimen2}567
	 \advance\count2 by-\ht2 \advance\count2 by-\dp2
	 \if.#3.\setbox4=\hbox to\wd0{\borderelm3\hss\borderelm4}
	 \else\setbox4=\hplusborder{\dimen2}384
	 \fi
	 \dimen0=\ht4\advance\dimen0 by\dp4
	 \advance\count2 by\dimen0\advance\count2 by-100
	 \divide\count2 by\dimen0
	 \box0
	 \cleaders\copy4\vskip\count2\dimen0
	 \box2
	 \kern0sp
	}%
   }

\ifx\manfnt\UNDEFINED
\font\manfnt=manfnt
\fi

\def\roundcorners#1
   {{\manfnt
     \setbox2=\hbox{\char'44}
     \ifcase#1
	\setbox0=\hbox{d}
	\hbox to\ht0{\vbox to\fontdimen8\font{\vss\box0}\hss}%
     \or\hbox{\vrule width\wd2 height\fontdimen8\font depth0pt\hss}
     \or\setbox0=\hbox{a}
	\hbox to\ht0{\kern\dp0\vbox to\fontdimen8\font{\vss\box0}\hss}%
     \or\setbox0=\hbox{d}%
	\hbox to\ht0{\vrule width\fontdimen8\font height\ht2 depth\dp2\hss}
     \or\setbox0=\hbox{d}%
	\hbox to\ht0{\hss\vrule width\fontdimen8\font height\ht2 depth\dp2}
     \or\setbox0=\hbox{c}
	\hbox to\ht0{\vbox{\copy0\kern-\dp0}\hss}%
     \or\hbox{\vrule width\wd2 height\fontdimen8\font depth0pt\hss}
     \or\setbox0=\hbox{b}
	\hbox to\ht0{\kern\dp0\vbox{\copy0\kern-\dp0}\hss}%
     \or\box2 
     \fi
   }}

\newbox\brd_txt
\newbox\brd_box

\def\setborder[#1,#2,#3]#4
   {\def\brd_inbox
       {\hsize=#1
	\advance\hsize by-#2
	\advance\hsize by-#2
	\aftergroup\set_border
       }%
    \def\set_border
       {\setbox\brd_txt=\vbox
	 {\vskip#3
	  \box\brd_txt
	  \vskip#3
	 }%
	\setbox\brd_box=#4{\ht\brd_txt}{#1}%
	\vbox to\ht\brd_box
	 {\copy\brd_box \kern-\ht\brd_box
	  \vss
	  \hbox to\wd\brd_box{\hss\box\brd_txt\hss}
	  \vss
	 }%
       }%
    \afterassignment\brd_inbox
    \setbox\brd_txt=\vbox
   }

\catcode`\_=\borderelm 

\let\borderelm=\roundcorners 

\newcommand{\drv}[2]{\frac{d#1}{d#2}}
\def\stacking#1#2#3{\mathrel{\mathop{#3}\limits^{#1}\limits_{#2}}}

\begin{document}
\title{A Theory for the Membrane Potential of Living Cells}
\author{L.P. Endresen $\, \dagger$ K. Hall $\, \ddagger$ J.S. H{\o}ye
$\, \dagger$ and J. Myrheim $\, \dagger$} \address{$\dagger
\,$Institutt for fysikk, NTNU, N-7034 Trondheim, Norway and $\ddagger
\,$Centre for Nonlinear Dynamics, McGill University, Montreal, Canada}
\date{\today} \maketitle

\begin{abstract} 
We give an explicit formula for the membrane potential of cells in
terms of the intracellular and extracellular ionic concentrations, and
derive equations for the ionic currents that flow through channels,
exchangers and electrogenic pumps.  We demonstrate that the work done
by the pumps equals the change in potential energy of the cell, plus
the energy lost in downhill ionic fluxes through the channels and
exchangers. The theory is illustrated in a simple model of
spontaneously active cells in the cardiac pacemaker.  The model
predicts the experimentally observed intracellular ionic concentration
of potassium, calcium, and sodium. Likewise the shapes of the
simulated action potential and five membrane currents are in good
agreement with experiments. We do not see any drift in the values of
the concentrations in a long time simulation, and we obtain the same
asymptotic values when starting from the full equilibrium situation
with equal intracellular and extracellular ionic concentrations.
\end{abstract} 

\section{Introduction} 
The purpose of the work we present here is to obtain a model for the
membrane potential of a single cell which is reasonably realistic, and
yet so simple that it can be used in practice to simulate numerically
single cells or several coupled cells. For simplicity, experimentally
observed currents (Boyett et al. 1993)  are omitted if they
seem too small to have a significant influence on the intracellular
ion concentration, or at least too small to change the dynamics of the
cell. On the other hand, we try to make the theory realistic by using
equations that are compatible with, or can be derived from, basic
physical principles.

It is a basic assumption of most models (Wilders 1993)  for the
electrical activity of cells that only the motion of positive ions,
and specifically those of potassium, calcium and sodium, influence the
membrane potential. This assumption is usually expressed as a
differential equation for the time dependence of the potential. We
observe that this differential equation can be integrated exactly, and
argue that the integration constant is given by the requirement that
the potential is zero when the ion concentrations on both sides of the
membrane are equal, as the density of negative charge happens to be
the same on both sides. Then it follows that the potential is directly
proportional to the excess number of positive ions inside the cell, a
formula which is nothing but the one for an electric capacitance that
follows from Gauss's law in electrostatics. 

We derive equations for ionic currents flowing through channels,
exchangers and electrogenic pumps. These are based on the Boltzmann
distribution law (Boltzmann 1868), which states that a particle in
thermal equilibrium spends less time in states of higher energy than
in states of lower energy, the Markov assumption (Markov 1906) which
says that the transition probabilities of a stochastic system (of
Markov type) is only dependent on its present state, and the principle
of detailed balance (Onsager 1931) which says that the microscopic
laws of physics are invariant with respect to the reversal of
time. Our equations were inspired by Ehrenstein and Lecar's model of
channel gating (1977), Nonner and Eisenberg's model for channel
current (1998), Mullins' model of the ${\rm Na}^{+},{\rm Ca}^{2+}$
exchanger (1977), and Chapman's model of the ${\rm Na}^{+},{\rm
K}^{+}$ pump (1978). In particular the book of Lorin John Mullins
(1981) ``Ion Transport in Heart'' has been a major source of
inspiration for us.

The theory is illustrated with a simple model of spontaneously active
cells in the rabbit sinoatrial node. The observable parameters in the
model are based on the experiments of Shibasaki (1987), Hagiwara {\rm
et al.} (1988), Muramatsu {\rm et al.} (1996) and Sakai {\rm et al.}
(1996). The non--observable parameters in the model are determined
numerically, in the same way as in an earlier study (Endresen 1997a),
by comparing the action potentials generated by the model with the
shape of the action potentials recorded by Baruscotti {\rm et al.}
(1996).

By using an algebraic equation for the potential in place of the
standard differential equation, as mentioned above, we obtain a model
which is stable against a slow drift of the intracellular ion
concentrations, sometimes seen in other models. Furthermore, by fixing
the integration constant for the voltage we obtain from the model a
prediction of the steady state ion concentrations in the cell. It is
even possible to predict these steady state concentrations by starting
with an initial state having equal concentrations inside and outside
the cell, and integrating the equations of motion over a long time
interval. 

{From} the equations of motion we obtain an equation that explicitly
demonstrates the energy balance in the process of moving ions in and
out of the cell.  The energy to make the cell function comes from
breakdown of ATP that runs the ${\rm Na}^{+},{\rm K}^{+}$ pump. Part
of this free (or useful) energy is dissipated while the rest enters
the cell. In the cell some of this energy is used to create a
potential energy that depends upon the ionic concentrations in the
cell, while the rest is dissipated by the currents in the ionic
channels and the ${\rm Na}^{+},{\rm Ca}^{2+}$ exchanger. This
potential energy function is thus such that the work associated with
ionic currents balances exactly the change in potential energy. In a
numerical integration of the differential equations one may compute
separately the work and potential energy, comparing the two in order
to check (and maybe control) the accuracy of the numerical
integration. In our long time integration we observe indeed a balance
between work and change in cell membrane potential energy.

\section{Derivation of the Equations}
\subsection{The Nernst Equilibrium Potential}
There are two basic principles behind the average motion of
particles. The first is diffusion, which is general; the second applies
only to charged particles such as ions in solutions. Simple diffusion
is described by the empirical law of Fick (1855),
\begin{equation}
\label{eq1}
\vec{\phi} = - ukT \nabla [{\rm S}] \;,
\end{equation}

\noindent
where ${\phi}$ is the ionic flux, $[{\rm S}]$ the concentration of
ions and $u$ the ratio of the velocity to the force acting on a
particle, known as the mobility. $T$ is the absolute temperature and
$k$ is Boltzmann's constant. The empirical law of Ohm (1827) describes
the net motion of charged particles in an electric field,
\begin{equation} 
\label{eq2} 
\vec{\phi} = - zeu [{\rm S}] \nabla U \;,
\end{equation}

\noindent
where $z$ is the valence, $e$ the elementary charge and $U$ the
electrical potential. Since we assume that the temperature is
constant, we can neglect the thermal flux given by Fourier's empirical
law. The fact that the mobility in Fick's law must be identical to the
mobility in Ohm's law was first noticed by Einstein (1905). If we
combine Eqs.~(\ref{eq1}) and (\ref{eq2}), the total flux of ions
due to diffusion and electric forces is
\begin{equation}
\label{eq3}
\vec{\phi} = - ukT \exp\left(-\frac{zeU}{kT}\right) \nabla \left[[{\rm S}] \exp\left(\frac{zeU}{kT}\right)\right] \;.
\end{equation}

\noindent
The equilibrium potential for which the flux is zero, is 
\begin{equation}
\label{eq4}
v_{\rm S} = U_{\rm i}-U_{\rm e} =  \frac{kT}{ze} \ln \left( \frac{[{\rm S}]_{\rm e}}{[{\rm S}]_{\rm i}}\right)\;.
\end{equation}

\noindent
It can be found by
setting $\vec{\phi}=0$ in Eq.~(\ref{eq3}) and integrating from
the extracellular (e) to the intracellular (i) side of the membrane.
Here $U_{\rm i}$, $U_{\rm e}$,
$[{\rm S}]_{\rm i}$ and $[{\rm S}]_{\rm e}$ are the
intracellular and extracellular potentials and concentrations.
This equation, first
stated by Nernst (1888) is based only on the empirical laws of Ohm and
Fick and the relation of Einstein.

The same formula can be derived in
a more general way using the Boltzmann factor (Boltzmann 1868). The
relative probability at equilibrium that an ion is at the
intracellular or extracellular side of a cell membrane is
\begin{equation}
\label{eq5}
\frac{p_{\rm i}}{p_{\rm e}} = \frac{[{\rm S}]_{\rm i}}{[{\rm S}]_{\rm e}} = \exp\left(-\frac{ze(U_{\rm i} - U_{\rm e})}{kT}\right) \;,
\end{equation}

\noindent
where $ze(U_{\rm i} - U_{\rm e})$ is the energy difference between the
two positions of the ion. Solving (\ref{eq5}) for $U_{\rm i} - U_{\rm
e}$ gives (\ref{eq4}). With the definition 
\begin{equation}
v_T=\frac{kT}{e}=\frac{RT}{F}\;,
\end{equation}
the equilibrium
potentials for the predominant cellular cations are then

\begin{eqnarray}
\label{eq7}
v_{\rm K}  &=& v_T \ln \frac{[{\rm K}]_{\rm e}}{[{\rm K}]_{\rm i}} \;, \\
\label{eq8}
v_{\rm Ca} &=& \frac{v_T}{2} \ln \frac{[{\rm Ca}]_{\rm e}}{[{\rm Ca}]_{\rm i}} \;, \\
\label{eq9}
v_{\rm Na} &=& v_T \ln \frac{[{\rm Na}]_{\rm e}}{[{\rm Na}]_{\rm i}} \;.
\end{eqnarray}

\subsection{Ionic Channels}
\subsubsection{Ionic Channel Gating}
Imagine that ionic channels are either completely open or completely
closed and randomly fluctuate between these states in a simple Markov
process (Markov 1906), described by the first order kinetics
(Ehrenstein and Lecar 1977)
\begin{equation}
\label{eq10}
\begin{array}{ccc}
{\displaystyle C} &
\stacking{\alpha}{\beta}{\displaystyle \rightleftharpoons}
& {\displaystyle O}
\end{array} \;,
\end{equation}
where the rate constants ${\alpha}$ and ${\beta}$ are functions of
transmembrane voltage and control the transitions between the closed
($C$) and the open ($O$) states of the gate. The rate for a closed
channel to open is ${\alpha}$, and ${\beta}$ is the rate for an open
channel to close. Let $x$ denote the average fraction of channels that
are open, or, equivalently, the probability that a given channel will
be open. We may say that the ionic flux through an ensemble of
channels is regulated by a sliding door whose position is $x$. This
yields:
\begin{equation}
\label{eq11}
\drv{x}{t} = \alpha (1-x) - \beta x \equiv \frac{x_{\infty} -x}{{\tau}} \;,
\end{equation}
where
\begin{eqnarray}
\label{eq12}
x_{\infty} &=& \frac{{\alpha}}{{\alpha} + {\beta}} \\
\label{eq13}
\tau   &=& \frac{1}{{\alpha} + {\beta}}\;.
\end{eqnarray}

\noindent
Here $x_{\infty}$ denotes the steady state fraction of open channels
and ${\tau}$ the relaxation time. Let us assume that the energy
difference between the open and closed positions is given by

\begin{equation}
\label{eq14}
\Delta G =  G_{\rm open} - G_{\rm closed} \equiv q (v_{\rm x}-v)\;,
\end{equation}

\noindent 
where $q$ is a gating charge, usually $q \approx \pm 4e$, such that $qv$
represents the change in electrical potential energy due to the
redistribution of charge during the transition, and where the term
$qv_{\rm x}$ represents the difference in mechanical conformational
energy between the two states. At equilibrium, $dx/dt=0$, and the
ratio of the probabilities for a single channel to be in the open
state or the closed state is
\begin{equation}
\label{eq15}
\frac{x_{\infty}}{1-x_{\infty}} = \frac{{\alpha}}{{\beta}}\;.
\end{equation}

\noindent
This relation is known as the principle of detailed balance (Onsager,
1931). The same ratio is given by the Boltzmann distribution
(Boltzmann 1868),
\begin{equation}
\label{eq16}
\frac{x_{\infty}}{1-x_{\infty}} = \exp\left(-\frac{\Delta G}{kT}\right) \;.
\end{equation}

\noindent
Thus, from Eqs.~(\ref{eq14}), (\ref{eq15}), and (\ref{eq16}), with $q=+4e$

\begin{equation}
\label{eq17}
{x}_{\infty} = \left[ {1+\exp\left(\frac{4e(v_{\rm x}-v)}{kT}\right)}\right]^{-1} \;.
\end{equation}

\noindent
The simplest possible choice for $\alpha$ and $\beta$ is

\begin{eqnarray}
\label{eq19}
{\alpha} &=& \lambda \exp\left(-\frac{2e(v_{\rm x}-v)}{kT}\right) \\
\label{eq20}
{\beta} &=& \lambda \exp\left(+\frac{2e(v_{\rm x}-v)}{kT}\right)\;,
\end{eqnarray}

\noindent
\noindent
where $\lambda$ is a constant. Assuming $\lambda$ to be constant gives
the maximum relaxation time at the voltage where $x_{\infty}=1/2$. The
relaxation time as a function of $v$ is then

\begin{equation}
\label{eq21}
{\tau} = \frac{1}{{\alpha}+{\beta}} = {\left[2\lambda \cosh\left(\frac{2e(v_{\rm x}-v)}{kT}\right) \right]}^{-1}\;.
\end{equation}

\subsubsection{Ion Channel Current}

Here we want to obtain the current $i$ through a one--dimensional
ionic channel of length $d$. We will allow the cross sectional area
$A$ to vary with position, i.e., we take $A=A(x)$.  By definition,
$x=-d/2$ is the inside and $x=d/2$ the outside of the membrane.  Let
$\phi=\phi(x)$ be the $x$-component of the flux $\vec{\phi}$, the
other components are negligible as long as the variation of $A$ with
$x$ is smooth and slow.  This is the analogue of water flow in a pipe
of varying cross section. By stationary flow, the current $i$ must be
the same through all cross sections, i.e. independent of $x$. Thus the
flux $\phi$ is inversely proportional to the area $A$, by the relation
\begin{equation}
\label{eq22}
i = ze \phi A = {\rm const}.
\end{equation}

\noindent
We insert $\phi$ from this equation in the
$x$ component of Eq.~(\ref{eq3}),
and multiply the resulting equation by
$\exp\left(ze(U-U_0)/kT\right)$,
introducing a constant voltage $U_0$ chosen such that
\begin{equation}
\label{eq25} 
U\!\left(-\frac{d}{2}\right) = U_0+\frac{v}{2} \;, \qquad 
U\!\left(\frac{d}{2}\right) = U_0-\frac{v}{2} \;.
\end{equation} 

\noindent
Then we find that
\begin{equation}
\label{eq23}
\frac{i}{A}\,\exp\left(\frac{ze(U-U_0)}{kT}\right) =
- zeukT \frac{d}{dx} 
\left[[{\rm S}] \exp\left(\frac{ze(U-U_0)}{kT}\right)\right] \;.
\end{equation}

\noindent
Here $U$, $[{\rm S}]$ and $A$ are functions of $x$, while all other
quantities are constant. (Note however that the mobility $u$ may be
reduced in a very narrow channel; one possible way to take into
account such an $x$ dependence of $u$ is to replace the area $A$ by an
effective area $A_{\rm eff}$ which is smaller than $A$).  Integrating
from the inside $x=-d/2$ to the outside $x=d/2$ we obtain
\begin{equation}
\label{eq24}
i = - \frac{zeukT}{I}\,
\left[[{\rm S}]_{\rm e} \exp\left(-\frac{zev}{2kT}\right) - 
      [{\rm S}]_{\rm i} \exp\left(\frac{zev}{2kT}\right) \right] \;,
\end{equation}

\noindent
where
\begin{equation}
\label{eq24b}
I= \int_{-d/2}^{d/2}
\frac{1}{A}\,\exp\left(\frac{ze(U-U_0)}{kT}\right)dx\;.
\end{equation}

\noindent
The concentrations are $[{\rm S}]_{\rm i}$ on the inside and
$[{\rm S}]_{\rm e}$ on the outside. If we extract a factor
$\sqrt{[{\rm S}]_{\rm e} [{\rm S}]_{\rm i}}$, and write the ratio of
the concentrations in terms of the Nernst potential defined in
Eq.~(\ref{eq4}), Eq.~(\ref{eq24}) can be written in
the following way,

\begin{eqnarray}
\label{eq26}
i &=&\frac{zeukT}{I}\;
\sqrt{[{\rm S}]_{\rm e} [{\rm S}]_{\rm i}} \left[
\sqrt{\frac{[{\rm S}]_{\rm i}}{[{\rm S}]_{\rm e}}}
\exp\left(\frac{zev}{2kT}\right) -
\sqrt{\frac{[{\rm S}]_{\rm e}}{[{\rm S}]_{\rm i}}}
\exp\left(-\frac{zev}{2kT}\right) \right] \nonumber \\ &=&
\frac{2zeukT}{I}\;
 \sqrt{[{\rm S}]_{\rm e} [{\rm S}]_{\rm i}}
\sinh\left(\frac{ze(v-v_{\rm S})}{2kT}\right) \;.
\end{eqnarray}

\noindent
Eq.~(\ref{eq26}) is our general result that follows from the
combined Ohm's and Fick's law.

Now the integral $I$ depends upon both the voltage $U=U(x)$ and the
cross section $A=A(x)$. To determine the $x$ dependence of $U(x)$ one
would need Poisson's equation for the electrostatic potential, taking
into account the net charge distribution in the membrane, including
both positive and negative ions.  However, this charge distribution
will depend upon detailed properties of membranes and their channels
that have been little known so far.  Thus it seems a reasonable
approach to make certain assumptions directly about $U(x)$.

A commonly used assumption is that $U(x)$ is linear, i.e.\ that
the electric field $-dU/dx$ is constant, and
that the cross section is constant, $A(x) = A_0$. Then Eq.
(\ref{eq26}) takes the form

\begin{equation}
\label{eq27}
i = (ze)^2 u\,\sqrt{[{\rm S}]_{\rm e} [{\rm S}]_{\rm i}}\;
\frac{A_0v\sinh\left(\frac{ze(v-v_{\rm S})}{2kT}\right)}
{d\sinh\left(\frac{zev}{2kT}\right)} \;.
\end{equation}

\noindent
As should be expected, this relation simplifies to the usual Ohm's law
in the special case $[{\rm S}]_{\rm i}=[{\rm S}]_{\rm e}$, since then
$v_{\rm S} = 0$ by Eq.~(\ref{eq4}). Eq.~(\ref{eq27}) is
known as the Goldman constant field approximation. Goldman (1943)
wrote:
\begin{quote}
{\footnotesize We assume that the membrane contains a large number of
dipolar ions near the isotonic point and that these can act to
minimize distortion in the field especially at low currents. We then
approach a situation in which the field is constant and are led to a
solution analogous to that given by Mott (1939) for electronic
conduction in the copper--copper oxide rectifier.}
\end{quote}

A more general case, perhaps somewhat more realistic, where the
integral $I$ can still be calculated exactly, is that of an ion
channel having a constant area $A_0$, except for a short and narrow
constriction or pore in its middle, with an area $A_p$ much smaller
than $A_0$.  An example is a cylindrical pore of radius $3 \, {\rm
\AA}$ and length $5 \, {\rm \AA}$, which is typical for ionic
channels. If we furthermore assume a constant electric field
everywhere in the channel, and if the length of the pore is $\epsilon
d$, then we have that
\begin{equation}
I=\frac{2dkT}{zev}\left[
\frac{1}{A_0}\sinh\left(\frac{zev}{2kT}\right)+
\left(\frac{1}{A_p}-\frac{1}{A_0}\right)
\sinh\left(\frac{\epsilon zev}{2kT}\right)
\right]\;.
\end{equation}
The limit of this as $v\to 0$ is
\begin{equation}
I_0=d\left[\frac{1-\epsilon}{A_0}+\frac{\epsilon}{A_p}\right]
\approx \frac{\epsilon d}{A_p}\;.
\end{equation}
The last approximation holds when the contribution from the pore
dominates the integral, which will be the case e.g.\ when the ratio of
areas, $A_p/A_0$, is of the order $\epsilon^2$.  For $\epsilon v$
small but nonzero the $v$ dependence of $I$ is only of second order in
$\epsilon v$, thus it will be a good approximation in a finite voltage
range to take $I$ to be constant, equal to $I_0$. The approximation
$I=\;\,$constant which is also valid under more general conditions
than those assumed in the above oversimplified derivation, and it
gives
\begin{equation}
\label{eq29}
i = k_{\rm S} \; \sinh\left(\frac{ze(v-v_{\rm S})}{2kT}\right) \;.
\end{equation}

\noindent
Here $k_{\rm S}$ is independent of $v$, e.g.\ in the case
considered above,

\begin{equation}
\label{eq30}
k_{\rm S} = 2zeukT\;\sqrt{[{\rm S}]_{\rm e} [{\rm S}]_{\rm i}}\;
\frac{ A_{\rm p}}{\epsilon d} \;.
\end{equation}

For Na and K ions it is a good approximation to consider the square
root of the concentrations $\sqrt{[{\rm S}]_{\rm e} [{\rm S}]_{\rm
i}}$ constant, while for Ca ions the relative change in concentration
is more significant during one action potential. In the present work
we used $k_{\rm S} = \;\,$constant in all three cases, for
simplicity. We have checked that this
does not affect our numerical results significantly.

There is reason to ask whether the linear voltage profile $U(x)$ can
be a reasonable approximation in the presence of a pore. Indeed, it
might seem natural to conclude that most of the voltage drop must be
concentrated at the pore due to its large resistance.  However, with
the combined Ohm's and Fick's law, the current is driven by gradients
in both voltage and concentration, as follows from Eq.~(\ref{eq3}).  A
large current may be due to a large voltage drop over the pore, or it
may be due to a large change in concentration, without any large
voltage difference. Thus, in general one has to take into account the
detailed properties of the channel in order to see which one of the
gradients is the dominant driving force in a given situation.

In a recent investigation by Nonner and Eisenberg (1998), Poisson's
equation relating the net charge density and electrostatic
potential was included in a more extensive analysis for a specific
model of a channel with a narrow pore. In their analysis they indeed
find that only part of the voltage drop is across the pore (something
like half of it). In their numerical simulations the voltage in the
pore is dominated by the presence of charged carboxyl groups, and thus
almost independent of the transmembrane voltage.  This lends support
to the approximation used here, that the integral $I$,
Eq.~(\ref{eq24b}), can be regarded as being constant.

Thus our simple result (\ref{eq29}) has the characteristic features of
the current--voltage relationships obtained by Nonner and Eisenberg in
their more extensive investigation. One characteristic feature is that
Eq.~(\ref{eq29}) shows inward rectification for large values of $[{\rm
S}]_{\rm e}/[{\rm S}]_{\rm i}$, i.e. increased conductance for large
negative potentials. Indeed the curves in figure 3A in Nonner and
Eisenberg (1998) have shapes of a hyperbolic sine. Such a behavior is
not predicted by the Goldman (1943) equation, Eq.~(\ref{eq27}), but
is seen in many excitable cells (Hille 1992). This is another reason
to base our computations on Eq.~(\ref{eq29}) in order to see the
consequences of its application.

\subsubsection{Potassium Channels}

If the flux of ions is given by Eq.~(\ref{eq29}) and regulated
by the fraction of open channels $x$, the membrane current through
potassium channels is 

\parbox{14cm}{
\begin{eqnarray*} 
i_{\rm K}    &=& k_{\rm K}\,x
\sinh\left(\frac{e(v-v_{\rm K})}{2kT}\right) \\
\frac{dx}{dt} &=& \frac{1}{\tau_{\rm K}}
\cosh\left(\frac{2e(v-v_{\rm x})}{kT}\right)
\left\{\frac{1}{2}\left[1+
\tanh\left(\frac{2e(v-v_{\rm x})}{kT}\right)\right]-x\right\} \;,
\end{eqnarray*}} \hfill
\parbox{1cm}{\begin{equation}\label{eq31}
\end{equation}}

\noindent
where ${\tau}_{\rm K} = 1/{2\lambda}$ is the maximum value of the
relaxation time, $k_{\rm K}$ is the conductance parameter of Eq.
(\ref{eq30}), $v_{\rm K}$ is given by Eq.~(\ref{eq7}), and the time
dependence of $x$ is given by Eq.~(\ref{eq11}) with Eqs.~(\ref{eq17})
and (\ref{eq21}) for $x_{\infty}$ and $\tau$ respectively. Here we
have used the identity
\begin{equation}
\label{eq32}
\frac{1}{2} \left(1+ \tanh\phi\right) = \frac{1}{1+\exp(-2\phi)} \;.
\end{equation}

\subsubsection{Calcium and Sodium Channels}

The calcium and sodium channels have an inactivation mechanism in
addition to the above activation mechanism. We can view these
mechanisms as two independent Markov processes, or equivalently two
independent sliding doors, which are both affected by voltage. An ion
can only go through if both sliding doors are at least slightly
open. Here the activation mechanism is very fast, with a time constant
of only a few milliseconds, so we use the steady state fraction of
open channels, Eq.~(\ref{eq17}), for this. The maximum time constant
of inactivation for calcium and sodium channels are of the same order
of magnitude as the maximum time constant of the activation of the
potassium channel (typically a few hundred milliseconds), thus 

\parbox{14cm}{
\begin{eqnarray*}
i_{\rm Ca} &=& k_{\rm Ca}\,f\,d_{\infty}
\sinh\left(\frac{e(v-v_{\rm Ca})}{kT}\right) \\
  d_{\infty} &=& \frac{1}{2}\left[1+
\tanh\left(\frac{2e(v-v_{\rm d})}{kT}\right)\right] \\
\frac{df}{dt} &=& \frac{1}{\tau_{\rm Ca}}
\cosh\left(\frac{2e(v-v_{\rm f})}{kT}\right)\left\{\frac{1}{2}\left[1-
\tanh\left(\frac{2e(v-v_{\rm f})}{kT}\right)\right]-f\right\} \;, \\
\end{eqnarray*}} \hfill
\parbox{1cm}{\begin{eqnarray}\label{eq33}\end{eqnarray}}

\noindent
and,

\parbox{14cm}{
\begin{eqnarray*}
i_{\rm Na}    &=& k_{\rm Na}\,h \,m_{\infty}
\sinh\left(\frac{e(v-v_{\rm Na})}{2kT}\right) \\
   m_{\infty} &=& \frac{1}{2}\left[1+
\tanh\left(\frac{2e(v-v_{\rm m})}{kT}\right)\right] \\
\frac{dh}{dt} &=& \frac{1}{\tau_{\rm Na}}
\cosh\left(\frac{2e(v-v_{\rm h})}{kT}\right)\left\{\frac{1}{2}\left[1-
\tanh\left(\frac{2e(v-v_{\rm h})}{kT}\right)\right]-h\right\} \;,
\end{eqnarray*}} \hfill
\parbox{1cm}{\begin{eqnarray}\label{eq34}\end{eqnarray}}

\noindent
where $k_{\rm Ca}$ and $k_{\rm Na}$ are the conductance parameters of
the calcium and sodium currents respectively, $v_{\rm Ca}$ and $v_{\rm
Na}$ are given by Eqs.~(\ref{eq8}) and (\ref{eq9}), $v_{\rm d}$ and
$v_{\rm m}$ are the half--activation potentials, and $v_{\rm f}$ and
$v_{\rm h}$ are the half--inactivation potentials. 

Note that the activation and inactivation mechanisms work in the same
way, and differ in two respects only. First, the time constants differ
experimentally by two orders of magnitude, and second, the gating
charge $q$, Eq.~(\ref{eq14}), is $+4e$ in one case and $-4e$ in the
other case.

\subsection{${\rm Na}^{+},{\rm K}^{+}$ Pump}

The Na,K--ATPase is found in the plasma membrane of virtually all
animal cells and is responsible for active transport of sodium and
potassium. Low sodium concentration and high potassium concentration
in the cytosol are essential for basic cellular functions such as
excitability, secondary active transport, and volume regulation. In
our model, the ${\rm Na}^{+},{\rm K}^{+}$ pump is the only energy
source.  We shall assume that the following equation is a complete
macroscopic description of the pump reaction (Chapman 1978),

\begin{equation}
\label{eq35}
\begin{array}{ccc}
{\rm ATP} + 3 {\rm Na}_{\rm i}^{+} + 2 {\rm K}_{\rm e}^{+} & \stacking{{\LARGE \alpha}}{{\LARGE \beta}} {\rightleftharpoons} & {\rm ADP} + {\rm P}_{\rm io} + 3 {\rm Na}_{\rm e}^{+} + 2 {\rm K}_{\rm i}^{+} 
\end{array} \;,
\end{equation}

\noindent
where ATP, ADP and ${\rm P}_{\rm io}$ are adenosine triphosphate,
adenosine diphosophate and inorganic phosphate, while $\alpha$ and
$\beta$ are the rates for the forward and backward reactions. The
energy involved in the movement of 3 ${\rm Na}^{+}$ and 2 ${\rm
K}^{+}$ ions against their electrochemical gradients is

\begin{eqnarray}
\label{eq36}
{\Delta G}_{\rm Na} &=& - 3e (v-v_{\rm Na}) \\
\label{eq37}
{\Delta G}_{\rm  K} &=& + 2e (v-v_{\rm  K}) \;,
\end{eqnarray}

\noindent
where $v_{\rm K}$ and $v_{\rm Na}$ are given by Eqs.~(\ref{eq7}) and
(\ref{eq9}). This result is independent of the detailed interaction
between ions, molecules and the ATPase enzyme. Therefore, the total
change in Gibbs free energy  is
\begin{eqnarray}
\Delta G &=& {\Delta G}_{\rm ATP} + {\Delta G}_{\rm Na} +{\Delta G}_{\rm K} \nonumber \\
\label{eq38}
         &=& e(v_{\rm ATP} + 3v_{\rm Na} - 2v_{\rm  K} - v) \;,
\end{eqnarray}

\noindent 
where ${\Delta G}_{\rm ATP}$ is the energy associated with the
breakdown of ATP, and $v_{\rm ATP} = {\Delta G}_{\rm ATP}/e$.  Note
that $\Delta G$ has to be negative, at least when averaged over time,
but the sum ${\Delta G}_{\rm Na}+{\Delta G}_{\rm K}$ may very well be
positive, since ${\Delta G}_{\rm ATP}$ is large and negative. Thus,
part of the energy from ATP breakdown goes into increasing the free
energy of potassium and sodium ions, but much energy is dissipated,
since the energy available is actually much larger than the energy
required to translocate the potassium and sodium ions at small
negative membrane potentials.

In practice, such a pump or motorized swing door will quickly reach
saturation. We therefore choose the sum of the forward and backward
rates to be constant, resembling the maximum possible speed of the
swing door in the forward and backward directions,
\begin{equation}
\label{eq39}
\alpha + \beta = \lambda \;,
\end{equation}

\noindent
where $\lambda$ is a constant. At equilibrium, the forward 
reaction must occur just as frequently as the reverse reaction,
giving
\begin{equation}
\label{eq40}
\frac{\alpha}{\beta} = \exp\left(-\frac{\Delta G}{kT}\right) \;.
\end{equation}

\noindent
Solving Eqs.~(\ref{eq39}) and (\ref{eq40}) for $\alpha$ and
$\beta$ gives

\begin{eqnarray}
\label{eq41}
\alpha &=& \frac{\lambda \exp\left(-\frac{\Delta G}{kT}\right)}{1+\exp\left(-\frac{\Delta G}{kT}\right)} \\
\label{eq42}
\beta  &=& \frac{\lambda}{1+\exp\left(-\frac{\Delta G}{kT}\right)} \;.
\end{eqnarray}

\noindent
The difference 
\begin{equation}
\label{eq43}
\alpha-\beta = \lambda\;\frac{\exp\left(-\frac{\Delta G}{2kT}\right)
                             -\exp\left(\frac{\Delta G}{2kT}\right)}
                             {\exp\left(-\frac{\Delta G}{2kT}\right) 
                             +\exp\left(\frac{\Delta G}{2kT}\right)}
               = \lambda\tanh\left(-\frac{\Delta G}{2kT}\right) \;,
\end{equation}

\noindent
gives the net pump current for a cell with $M$ pumps as 
\begin{equation} 
\label{eq44}
i_{\rm NaK} = Me(\alpha-\beta) = k_{\rm NaK}
\tanh\left(\frac{e(v+2v_{\rm K}-3v_{\rm Na}-v_{\rm ATP})}{2kT}\right)\;,
\end{equation}

\noindent
where $k_{\rm NaK} = Me\lambda$.

\subsection{${\rm Na}^{+},{\rm Ca}^{2+}$ Exchanger}
To maintain a steady state for the intracellular calcium concentration
in for example heart cells, the amount of calcium that enters the cell
via ionic channels must be extruded. The ${\rm Na}^{+},{\rm Ca}^{2+}$
exchanger is the major mechanism responsible for achieving a balance
between calcium entry and extrusion in oscillating cells. We assume
that the rate for the forward ($\alpha$) and the backward ($\beta$)
exchange reaction given by (Mullins 1977)

\begin{equation}
\label{eq45}
\begin{array}{ccc}
3 {\rm Na}_{\rm e}^{+} + {\rm Ca}_{\rm i}^{2+} & \stacking{{\LARGE \alpha}}{{\LARGE \beta}} {\rightleftharpoons} & 3 {\rm Na}_{\rm i}^{+} + {\rm Ca}_{\rm e}^{2+} 
\end{array} \;,
\end{equation}

\noindent
are governed largely by the electrochemical gradients for sodium and
calcium, together with the membrane potential. In other words, the
energy produced when 3 extracellular sodium ions take the elevator
down into the cytosol is used to elevate one calcium ion up from the
cytosol into the extracellular space,
\begin{eqnarray}
\label{eq46}
{\Delta G}_{\rm Na} &=& + 3e (v-v_{\rm Na}) \\
\label{eq47}
{\Delta G}_{\rm Ca} &=& - 2e (v-v_{\rm Ca}) \;,
\end{eqnarray}

\noindent
where $v_{\rm Ca}$ and $v_{\rm Na}$ are given by Eqs.~(\ref{eq8})
and (\ref{eq9}). The total work done in reaction (\ref{eq45}) is
\begin{equation}
\label{eq48}
\Delta G = {\Delta G}_{\rm Na} +{\Delta G}_{\rm Ca} =  e (v - 3v_{\rm Na} + 2v_{\rm Ca}) \;.
\end{equation}

\noindent
The ratio of $\alpha$ to $\beta$ in Eq.~(\ref{eq45}) is again
determined by $\Delta G$ like in Eq.~(\ref{eq40}). However, in
the present case saturation effects are not expected and furthermore
$\Delta G$ will vary around zero, so we put 
\begin{eqnarray}
\label{eq49}
\alpha &=& \lambda
\exp\left(-\frac{e(v-3v_{\rm Na}+2v_{\rm Ca})}{2kT}\right) \\
\label{eq50}
\beta  &=& \lambda
\exp\left(+\frac{e(v-3v_{\rm Na}+2v_{\rm Ca})}{2kT}\right) \;,
\end{eqnarray}

\noindent
where we make the assumption that $\lambda$ is a constant (Mullins,
1981). For a cell with $N$ exchangers the net current is then 
\begin{equation}
\label{eq51}
i_{\rm NaCa} = -Ne(\alpha-\beta) 
             = k_{\rm NaCa}\,
\sinh\left(\frac{e(v- 3v_{\rm Na}+ 2v_{\rm Ca})}{2kT}\right) \;,
\end{equation}

\noindent
where $k_{\rm NaCa} = 2Ne\lambda$.

\subsection{Membrane Voltage}
Imagine that the electrical activity of a cell is described by the
five currents discussed above, and that all the other currents
(Boyett 1996) are of minor importance. The standard differential
equations for the voltage, and the conservation laws for intracellular
ionic concentrations are then
\begin{eqnarray}
\label{eq52}
\frac{dv}{dt} &=& -\frac{1}{C}
\left(i_{\rm K}+i_{\rm Ca}+i_{\rm Na}+i_{\rm NaCa}+i_{\rm NaK}
\right)\;,  \\
\label{eq53}
\frac{d}{dt}{[\rm K]_{\rm i}} &=& \frac {2i_{\rm NaK} - i_{\rm K}}{FV}\;, \\
\label{eq54}
\frac{d}{dt}{[\rm Ca]_{\rm i}} &=& \frac {2{i_{\rm NaCa}}- i_{\rm Ca}}{2FV}\;, \\
\label{eq55}
\frac{d}{dt}{[\rm Na]_{\rm i}} &=& \frac {-{i_{\rm Na}}- 3i_{\rm NaK}- 3i_{\rm NaCa}}{FV}\;,
\end{eqnarray}

\noindent
where $C$ is cell capacitance, $F$ is Faraday's constant, and we assume
that the cell volume $V$ is constant. Now Eqs.~(\ref{eq53}),
(\ref{eq54}) and (\ref{eq55}) can be solved for $i_{\rm K}$, $i_{\rm
Ca}$, and $i_{\rm Na}$, and we obtain
\begin{eqnarray}
\label{eq56}
i_{\rm K} &=& -FV \frac{d}{dt}{[\rm K]_{\rm i}} + 2i_{\rm NaK} \;, \\
\label{eq57}
i_{\rm Ca} &=& -2FV \frac{d}{dt}{[\rm Ca]_{\rm i}} + 2i_{\rm NaCa} \;, \\
\label{eq58}
i_{\rm Na} &=& -FV \frac{d}{dt}{[\rm Na]_{\rm i}} - 3i_{\rm NaK}- 3i_{\rm NaCa} \;.
\end{eqnarray}

\noindent
Inserting this into Eq.~(\ref{eq52}) yields
\begin{equation}
\label{eq59}
\frac{dv}{dt} = \frac{FV}{C}\,\frac{d}{dt}
\left({[\rm K]_{\rm i}}+2{[\rm Ca]_{\rm i}}+{[\rm Na]_{\rm i}} \right)\;,
\end{equation}

\noindent
since the remaining currents cancel. This equation can also be written
as

\begin{equation}
\label{eq60}
\frac{d}{dt} \left(v - \frac{FV}{C} \left\{ [\rm K]_{\rm i} + 2[\rm Ca]_{\rm i} + [\rm Na]_{\rm i}  \right\} \right) = 0\;.
\end{equation}

\noindent
This integrated gives
\begin{equation}
\label{eq61}
v - \frac{FV}{C}
\left( [\rm K]_{\rm i}+2[\rm Ca]_{\rm i}+[\rm Na]_{\rm i} \right) = v_0 \;,
\end{equation}
\noindent
where the integration constant $v_0$ has to be determined. Given that the
voltage across a capacitor is zero when the net charge difference is zero,
we must choose the integration constant
\begin{equation}
\label{eq62}
v_0 = -\frac{FV}{C} \left([\rm K]_{\rm e} + 2[\rm Ca]_{\rm e} + [\rm Na]_{\rm e} \right) \;,
\end{equation}
\noindent
which gives

\begin{equation}
\label{eq63}
v = \frac{FV}{C} \left\{ [\rm K]_{\rm i}-[\rm K]_{\rm e} + 2([\rm Ca]_{\rm i}-[\rm Ca]_{\rm e}) + [\rm Na]_{\rm i}-[\rm Na]_{\rm e} \right\} \;.
\end{equation}
\noindent
This choice of $v_0$ depends on the assumption that all other ions
have the same concentrations on both sides, consistent with Eq.
(\ref{eq52}) where it is assumed that they do not contribute to the
current. This is also consistent with standard assumptions in the
literature (Encyclop{\ae}dia Britannica 1997)

\begin{quote}
{\footnotesize In the extracellular fluid, electroneutrality is
preserved by a balance between a high concentration of ${\rm Na}^+$ on
the one hand and a high concentration of ${\rm Cl}^{-}$ as well as
small quantities of impermeant anions such as bicarbonate, phosphate,
and sulfate on the other. In the cytoplasm, where ${\rm K}^{+}$
concentration is high, the concentration of ${\rm Cl}^{-}$ is much
below that necessary to balance the sum of the positive
charges. Electroneutrality is maintained there by negatively charged
impermeant proteins and phosphates. Osmotic balance is maintained
between the extracellular fluid and the cytoplasm by movement of water
through the plasma membrane when the total concentration of particles
on one side is not equal to that on the other.}
\end{quote}

Eq.~(\ref{eq63}) is nothing but the relation between electric
potential and charge of a capacitor, which is actually
the origin of Eq.~(\ref{eq52}).  Thus it is completely general
and independent of the number of membrane currents in a model.  It
means that:
\begin{quote}
{\it The voltage across the membrane of a cell is caused by, and is
directly proportional to, the surplus of charge inside the cell.}
\end{quote}

\noindent
Since Eq.~(\ref{eq63}) is the explicit integral of Eq.  (\ref{eq52}),
it can be used instead of Eq.~(\ref{eq52}) (or the equivalent of
Eq.~(\ref{eq52})) in any model. The differential  equation,
Eq.~(\ref{eq52}), is needed only in models where the intracellular
ionic concentrations are not tracked individually (for example, the
Hodgkin--Huxley equations (1952)).

There is a significant difference between Eqs.~(\ref{eq52}) and
(\ref{eq63}) for use in numerical simulations, for the following
reason. There are two different ways to determine how many ions there
are inside a cell. The first method counts every ion entering or
leaving (Eq.~(\ref{eq52})), while the second method counts all
the ions inside the cell (Eq.~(\ref{eq63})). Both methods will
give correctly the {\em variation} in the number of ions in the
cell. However, the observer of ions entering and leaving observes only
the variations in the number, and if he wants to know the actual
number, he must make an initial guess of the number of ions already
inside. Because his guess may differ significantly from the actual ion
number, the results from the two methods may be contradictory.

A variant of Eq.~(\ref{eq63}) has recently been derived by
Varghese and Sell (1997). However, they did not identify the
integration constant $v_0$, which is related to the initial ionic
concentrations and represents the initial net charge via the electric
capacitance of the cell as shown in Eq.~(\ref{eq63}).

There is reason to ask whether it is a reasonable approximation to
omit the anions in Eq.~(\ref{eq63}). This can be justified if the
total concentration of cations is approximately the same on both
sides.  Indeed, this property is seen in most ionic models, like for
instance in Wilders (1993), where the cation concentrations are
\begin{equation}
\begin{array}{lll} 
 [{\rm K}]_{\rm e} = 5.4 \, {\rm mM} &
\quad [{\rm Ca}]_{\rm e} = 2  \, {\rm mM} &
\quad [{\rm Na}]_{\rm e} = 140  \, {\rm mM} \\
 {[{\rm K}]}_{\rm i} = 140  \, {\rm mM} &
\quad [{\rm Ca}]_{\rm i} = 0.0000804  \, {\rm mM}  &
\quad [{\rm Na}]_{\rm i} = 7.5  \, {\rm mM} \;.\\
\end{array} 
\end{equation} 

\subsection{Energy Balance and Osmotic Pressure}
\noindent

The current $i$ in Eq.~(\ref{eq22}) may be written as
\begin{equation}
i=ze\,\frac{dn}{dt}\;,
\end{equation}
where $n$ is the number of ions transferred from the inside to the
outside of the membrane, and $dn/dt$ is the rate of transfer. The
change in free energy when one ion is transferred, is $\Delta
G=-ze(v-v_S)$, and the total change in free energy over a time
interval is
\begin{equation}
\label{eq64a}
\Delta G
=-\int_0^n ze(v-v_S)\,dn
=-\int_0^t i\,(v-v_S)\,dt\;.
\end{equation}
It follows from Eq.~(\ref{eq29}) that the current $i$ has the same
sign as the voltage $v-v_S$ as required in general to have
thermodynamic stability, so that the integrand in Eq.~(\ref{eq64a}) is
positive (strictly speaking non-negative), and therefore $\Delta G<0$
(or $\Delta G\leq 0$).

By similar reasoning, taking into account all the reversal potentials
and the free energy associated with the breakdown of ATP, we find that
the total change in free energy due to the five currents in our model
is

\begin{eqnarray}
\label{eq65}
\Delta G &=& -\int_{0}^{t} \left[
 i_{\rm K}(v-v_{\rm K})
+i_{\rm Ca}(v-v_{\rm Ca})
+i_{\rm Na}(v-v_{\rm Na}) \right. \nonumber \\
&&+\left.
 i_{\rm NaCa}(v- 3v_{\rm Na}+2v_{\rm Ca})
+i_{\rm NaK}(v+2v_{\rm K}-3v_{\rm Na}-v_{\rm ATP})\right] dt\;.
\end{eqnarray}

\noindent
Each of the five terms in the integrand is positive (non-negative),
since each current has the same sign as the corresponding voltage. In
other words, energy is dissipated all the time by all the five
currents, implying that $\Delta G\leq 0$.

The main contribution to the negative $\Delta G$ is the ATP term,
\begin{eqnarray}
\label{eq65ATP}
\Delta G_{\rm ATP} = \int_{0}^{t} 
i_{\rm NaK}\,v_{\rm ATP}\,dt\;,
\end{eqnarray}

\noindent
which in practice is negative all the time, and furthermore is large
in magnitude compared to the other terms. It should be noted that this
term is the source of useful energy that is dissipated to maintain the
activity of the cell and keep it away from equilibrium. 
Keeping this term apart, we may calculate the change in free energy of
the ionic system,
\begin{eqnarray}
\label{eq65ions}
\Delta G_{\rm ions} &=& \Delta G-\Delta G_{\rm ATP} \nonumber\\
&=& -\int_{0}^{t} \left[
 i_{\rm K}(v-v_{\rm K})
+i_{\rm Ca}(v-v_{\rm Ca})
+i_{\rm Na}(v-v_{\rm Na}) \right. \nonumber \\
  &&+\left.
\label{eq70}
 i_{\rm NaCa}(v-3v_{\rm Na}+2v_{\rm Ca})
+i_{\rm NaK}(v+2v_{\rm K}-3v_{\rm Na}) \right] dt \;.
\end{eqnarray}

\noindent
Using Eqs.~(\ref{eq52}), (\ref{eq56}), (\ref{eq57}), and (\ref{eq58})
to eliminate the currents, and assuming the capacitance $C$ and the
volume $V$ to be constant, we get that
\begin{eqnarray}
\label{eq66}
\Delta G_{\rm ions} &=& C \int_{0}^{v} v\,dv
-FV \int_{[\rm K]_{\rm e}}^{[\rm K]_{\rm i}}
v_{\rm K}\,d([{\rm K}]_{\rm i})
\nonumber\\
&&-2FV \int_{[\rm Ca]_{\rm e}}^{[\rm Ca]_{\rm i}}
v_{\rm Ca}\,d([{\rm Ca}]_{\rm i})
-FV \int_{[\rm Na]_{\rm e}}^{[\rm Na]_{\rm i}}
v_{\rm Na}\,d([\rm Na]_{\rm i})\;,
\end{eqnarray}

\noindent
where the reversal potentials $v_{\rm K}$, $v_{\rm Ca}$, and $v_{\rm
Na}$ depend on the integration variables $[\rm K]_{\rm i}$, $[\rm
Ca]_{\rm i}$, and $[\rm Na]_{\rm i}$ according to Eqs.~(\ref{eq7}),
(\ref{eq8}) and (\ref{eq9}).  Integrating from the equilibrium state
$v=0$, $[\rm K]_{\rm i}=[\rm K]_{\rm e}$, $[\rm Ca]_{\rm i}=[\rm
Ca]_{\rm e}$, and $[\rm Na]_{\rm i}=[\rm Na]_{\rm e}$, using the
indefinite integral
\begin{equation}
\label{eq67}
\int \ln\phi\;d\phi = \phi \ln\phi - \phi \;,
\end{equation}

\noindent
we find

\parbox{14cm}{
\begin{eqnarray*}
\Delta G_{\rm ions} = \frac{1}{2}\,C v^2
&+& RTV \left\{
[\rm K]_{\rm i}\,\ln
\left({\frac {[\rm K]_{\rm i}}{[\rm K]_{\rm e}}}\right)
+[\rm Ca]_{\rm i}\,\ln
\left({\frac {[\rm Ca]_{\rm i}}{[\rm Ca]_{\rm e}}}\right)
+[\rm Na]_{\rm i}\,\ln
\left({\frac {[\rm Na]_{\rm i}}{[\rm Na]_{\rm e}}}\right)\right\} \\
&-& RTV\left(
 [\rm K]_{\rm i}-[\rm K]_{\rm e}
+[\rm Na]_{\rm i}-[\rm Na]_{\rm e}
+[\rm Ca]_{\rm i}-[\rm Ca]_{\rm e} \right) \,.
\end{eqnarray*}} \hfill
\parbox{1cm}{\begin{eqnarray}\label{eq68}\end{eqnarray}}

\noindent
In passing it may be noted that in the present case the equilibrium
state is simply the one with equal concentrations of cations on both
sides, as follows from our assumption of having the same concentration
of anions or negative charge on both sides of the membrane. More
generally equilibria for ionic systems are described by the Donnan
(1911) equilibrium that can yield different concentrations on both
sides of the membrane.

Since $\Delta G_{\rm ions}$ is a function only of the state of the
cell and is independent of the process by which the state is reached,
it represents a potential energy for the cell, which we will call $P$.
Note that $P=0$ in the equilibrium state, whereas $P>0$ in all other
states. $P$ is the minimum energy needed to bring a thermal
system away from equilibrium with its surroundings, or equivalently,
the maximum work that can be performed by the system when returning to
equilibrium.

The potential energy $P$, as defined in Eq.~(\ref{eq68}), contains
three terms, each of which can be given a more direct physical
interpretation. The first term is simply the electrostatic energy of a
capacitor, while the two temperature dependent terms are related to
thermal properties.
In fact, since we assume ideal dilute solutions, the change in entropy
due to changes of ion concentrations away from their equilibrium
values is
\begin{equation}
\label{eq78}
s = RV \left\{
[\rm K]_{\rm i}\,
\ln \left({\frac {[\rm K]_{\rm e}}{[\rm K]_{\rm i}}}\right)
+[\rm Ca]_{\rm i}\,
\ln \left({\frac{[\rm Ca]_{\rm e}}{[\rm Ca]_{\rm i}}}\right)
+[\rm Na]_{\rm i}\,
\ln\left({\frac {[\rm Na]_{\rm e}}{[\rm Na]_{\rm i}}}\right)\right\}\;.
\end{equation}

\noindent
Under the same changes, the change in osmotic pressure inside the cell
is equal to the difference in osmotic pressure across the membrane,
which is, for ideal solutions,
\begin{equation}
\label{eq79}
\pi = RT \left(
 [\rm K]_{\rm i}-[\rm K]_{\rm e}
+[\rm Na]_{\rm i}-[\rm Na]_{\rm e}
+[\rm Ca]_{\rm i}-[\rm Ca]_{\rm e} \right)\;.
\end{equation}
Note that for fixed volume the anions will not contribute to the
difference in osmotic pressure, but they will contribute if the volume
is changed and the membrane is impermeable to them.  In terms of the
change in transmembrane voltage, $v$, the change in entropy, $s$, and
the change in transmembrane osmotic pressure, $\pi$, as compared to
the equilibrium state, we may write
\begin{equation}
\label{eq77}
P = \frac{1}{2}\,Cv^2 - Ts - V\pi \;.
\end{equation}

Equation (\ref{eq79}) is the van't Hoff equation (1887) for the
osmotic pressure across a solute impermeable barrier separating two
ideal dilute solutions. In 1887 van't Hoff noticed that the behavior
of solutes in dilute solutions resembles the behavior of a perfect
gas (van't Hoff, 1887), and as quoted by Arrhenius in a memoir edited
by Jones (1899):

\begin{quote}
{\footnotesize The pressure which a gas exerts at a given temperature
if a definite number of molecules is contained in a definite volume,
is equal to the osmotic pressure which is produced by most substances
under the same conditions, if they are dissolved in any given liquid.}
\end{quote}

Rewriting Eq.~(\ref{eq68}), using Eqs.~(\ref{eq65ions}) and
(\ref{eq77}), we may summarize the energy balance in the following
way, 
\begin{eqnarray}
\label{eq80}
-\int_{0}^{t} i_{\rm NaK} (v + 2v_{\rm K} - 3v_{\rm Na})\,dt =
\frac{1}{2}\,C v^2-sT-\pi V
\phantom{(v-v_{\rm Na})+i_{\rm NaCa}(v-3v_{\rm Na}+2v_{\rm Ca}dt}
&&
\nonumber\\
+\int_{0}^{t}\left[
 i_{\rm K}(v-v_{\rm K})
+i_{\rm Ca}(v-v_{\rm Ca})
+i_{\rm Na}(v-v_{\rm Na})
+i_{\rm NaCa}(v-3v_{\rm Na}+2v_{\rm Ca})\right]dt&&.
\end{eqnarray}

\noindent
The left hand side of this equation is the useful work performed upon
the cell by the ${\rm Na}^{+},{\rm K}^{+}$ pumps, moving ${\rm
Na}^{+}$ and ${\rm K}^{+}$ ions against their potential gradients.
The energy supplied by the pumping of ions produces the following
effects that either change the potential energy of the cell or cause
energy loss by dissipation,
\begin{enumerate}
\item a transmembrane voltage difference, $v$;
\item a change in entropy, $s$;
\item a transmembrane osmotic pressure difference, $\pi$; and
\item downhill ionic currents through the exchangers and channels,
$i_{\rm K}$, $i_{\rm Ca}$, $i_{\rm Na}$, and $i_{\rm NaCa}$.
\end{enumerate}

In an oscillating cell, as described by the present model, the
following two inequalities will hold, over a sufficiently long time
interval,
\begin{equation}
\label{eq80aa}
-\Delta G_{\rm ATP}
= -\int_{0}^{t} i_{\rm NaK}\, v_{\rm ATP}\,dt
> -\int_{0}^{t} i_{\rm NaK}\,(v+2v_{\rm K}-3v_{\rm Na})\,dt
> 0\;.
\end{equation}
The first inequality is simply the inequality $i_{\rm NaK}(v+2v_{\rm
K}-3v_{\rm Na}-v_{\rm ATP})>0$, which follows from Eq.~(\ref{eq44}).
It means that the energy released by breakdown of ATP is larger than
the useful work performed by the pumps, as required from general
principles, in other words, that energy is dissipated by the current
$i_{\rm NaK}$ produced by the pumps. The second inequality must hold
due to Eq.~(\ref{eq80}), where the right hand side consists of three
oscillating potential energy terms plus four positive terms that
describe energy dissipation. This inequality shows that the useful
work performed by the pumps is positive, as required to maintain the
dissipation due to the other currents of a working cell away from
thermal equilibrium. 

We may remark that the laws of Ohm and Fick, equations (\ref{eq1}) and
(\ref{eq2}), are consistent with the use of ideal solutions and osmotic
pressure that assume independent (non--interacting) particles.
Arrhenius (1902) wrote about the relationship between osmotic pressure
and diffusion:
\begin{quote}
{\footnotesize Besides the electrical, other forces may be active in
causing the movement of the ions. Of these the osmotic pressure is the
most important. On account of this pressure a phenomenon called
diffusion (hydrodiffusion) may be observed.}
\end{quote}

\noindent
In the model considered the osmotic pressure $\pi$ has not been
involved in the dynamics. However in an extended model with variable
volume $V$ it will be more important as it will determine the solute
flux through the membrane.

\section{A Model for Cardiac Pacemaker Cells}
In the above, a mathematical model of the membrane potential has been
derived where Eqs.~(\ref{eq7}), (\ref{eq8}), and (\ref{eq9}) represent
the equilibrium potentials, Eqs.~(\ref{eq31}), (\ref{eq33}), and
(\ref{eq34}) the ionic currents, Eqs.~(\ref{eq51}) and (\ref{eq44})
the exchanger and the pump currents, Eqs.~(\ref{eq53}), (\ref{eq54})
and (\ref{eq55}) the ionic concentrations, Eq.~(\ref{eq63}) the
membrane voltage, and finally, Eq.~(\ref{eq77}) the osmotic pressure
across the cell membrane. The model has 6 time dependent variables
$x$, $f$, $h$, $[\rm K]_{\rm i}$, $[\rm Ca]_{\rm i}$ and $[\rm
Na]_{\rm i}$, and the equations are summarized in Appendix A.

\subsection{Ionic Mechanisms in the Cardiac Pacemaker}
Akinori Noma published in 1996 an excellent review of the ionic
mechanisms of the cardiac pacemaker potential (Noma, 1996). In this
short paper, Noma investigated the mechanisms that produce spontaneous
activity in sinoatrial node cells, and introduced the following
overview of the relevant ionic currents.

\subsubsection*{Channel gating which drives membrane depolarization during diastole}
\begin{itemize}
\item Deactivation of $i_{\rm K}$ ($i_{\rm Kr}$).
\item Removal of inactivation of $i_{\rm Ca,L}$ and $i_{\rm st}$.
\item Activation of the hyperpolarization--activated current ($i_{\rm f}$).
\item Activation of L--type ${\rm {Ca}^{2+}}$ current ($i_{\rm Ca,L}$).
\item Activation of T--type ${\rm {Ca}^{2+}}$ current ($i_{\rm Ca,T}$).
\end{itemize}

\subsubsection*{Background conductance}
\begin{description}
\item[$i_{\rm b,Na}$]: A cation current with reversal potential of about $-20 \, {\rm mV}$.
\item[$i_{\rm K,ACh}$]: Spontaneous openings of the ${\rm K}^{+}$ channels.
\item[$i_{\rm NaK}$]: Na/K pump current.
\item[$i_{\rm NaCa}$]: Na/Ca exchange current.
\item[$i_{\rm K,ATP}$]: ATP sensitive ${\rm K}^{+}$ channels.
\end{description}

We will not try to determine the relative amplitude of the above
current components here, but instead demonstrate that only five
membrane currents is sufficient to ensure stable intracellular ionic
concentrations. These are $i_{\rm K}$, $i_{\rm Ca}$, $i_{\rm NaK}$,
$i_{\rm NaCa}$, and $i_{\rm Na}$. In our model we assume that $i_{{\rm
Ca,T}}$ and $i_{{\rm Ks}}$ are of minor importance; i.e. when we talk
about $i_{\rm Ca}$ we mean $i_{{\rm Ca,L}}$, and when we talk about
$i_{\rm K}$ we mean $i_{{\rm Kr}}$.

\subsection{Model Parameters}
The various parameters play different roles in the model, and we list
them in different tables to distinguish between fundamental physical
constants (table \ref{table1}), experimentally observed constants
(table \ref{table2}), adjustable parameters (table \ref{table3}) and
initial conditions (table \ref{table4}) in the model. One parameter
not listed is the gating charge $q$, Eq.~(\ref{eq14}), which is the
origin of the factor $2e/kT$ in Eqs.~(\ref{eq31}), (\ref{eq33}) and
(\ref{eq34}). This corresponds to a slope factor for the activation
and inactivation curves of $kT/4e \approx 6.68 \, {\rm mV}$ at
$37^{\circ} {\rm C}$. The observed slope factors are $7.4 \, {\rm mV}$
for activation of $i_{\rm K}$ (Shibasaki 1987), $6.6 \, {\rm mV}$ for
activation of $i_{\rm Ca}$ (Hagiwara {\rm et al.} 1988), $6.0 \, {\rm
mV}$ for inactivation of $i_{\rm Ca}$ (Hagiwara {\rm et al.} 1988),
$6.0 \, {\rm mV}$ for activation of $i_{\rm Na}$ (Muramatsu {\rm et
al.}  1996), and, finally, $6.4 \, {\rm mV}$ for inactivation of
$i_{\rm Na}$ (Muramatsu {\rm et al.} 1996). Hence, we see that
$kT/4e$, corresponding to a gating charge of $q \approx \pm 4e$, is a
good approximation.

The half--activation and inactivation potentials in the model ($v_{\rm
x}$, $v_{\rm d}$, $v_{\rm f}$, $v_{\rm m}$ and $v_{\rm h}$) are based
on the experiments of Shibasaki (1987), Hagiwara {\rm et al.} (1988)
and Muramatsu {\rm et al.} (1996), and we use a value of $v_{\rm ATP}$
that gives a reversal potential for the sodium pump in good agreement
with the experiments of Sakai {\rm et al.} (1996). The maximum time
constants in these experiments were $ 203 \, {\rm ms}$ for activation
of $i_{\rm K}$ (Shibasaki 1987), $225 \, {\rm ms}$ for inactivation
of $i_{\rm Ca}$ (Hagiwara {\rm et al.} 1988) and $174 \, {\rm ms}$
for inactivation of $i_{\rm Na}$ (Muramatsu {\rm et al.} 1996). In
the model, however, we combine these and use a maximum time constant of
$200 \, {\rm ms}$ for both ${\tau}_{\rm K}$, ${\tau}_{\rm Ca}$ and
${\tau}_{\rm Na}$. Finally, we use typical values for cell volume, cell
capacitance, and extracellular ionic concentrations.

Two of the differential equations in the model have almost identical
but opposite dynamics. If we modify the half--inactivation potentials
of calcium from $v_{\rm f} = -25.0 \, {\rm mV}$ to $v_{\rm f} = v_{\rm
x} = -25.1 \, {\rm mV}$, it is possible to relate the inactivation
gating of calcium to the activation gating of potassium by
the equation

\begin{equation}
\label{eq85}
x + f = 1 \;,
\end{equation}

\noindent
since the time constants for these two processes are equal. We have
thus reduced the number of differential equations in the model by one.

This computational saving will be irrelevant for the computation of one action
potential, but is important in an extended model with thousands of
coupled cell, or in a long--time integration of the one cell model.

\subsection{Pacemaker Current}
The relative amplitude of the ionic currents that drive membrane
depolarization during diastole is still a matter of
debate. DiFrancesco (1993) argues that the hyperpolarization activated
current ($i_{\rm f}$) is the only current that can generate and
control the slow depolarization of pacemaker cells. The $i_{\rm f}$
current is normally carried by ${\rm Na}^{+}$ and ${\rm K}^{+}$.  Guo
{\rm et al.} (1995a; 1995b; 1996) reported another current, called the
sustained inward current $i_{\rm st}$, where the major charge carrier
is believed to be ${\rm Na}^{+}$. Also a ${\rm Ca}^{2+}$ ``window''
current has been observed in rabbit sinoatrial node cells (Denyer \&
Brown 1990). It is possible that any one of these currents, or a
combination of them, is responsible for membrane depolarization during
diastole. However, the estimates of the net membrane current during
diastole is so imprecise (Zaza {\rm et al.} 1997), that we could not
form a judgment on the question.

During diastole the electrochemical driving forces produce outward
${\rm K}^{+}$ currents and inward ${\rm Na}^{+}$ and ${\rm Ca}^{2+}$
currents, and the driving force for ${\rm Ca}^{2+}$ is much larger
than that for ${\rm Na}^{+}$. These findings implies that a
significant background influx of ${\rm Ca}^{2+}$ is possible during
diastole, and that this current might be responsible for the pacemaker
activity in sinoatrial node cells. We denote the conductance for this
current $k_{\rm b,Ca}$, and modify (\ref{eq33}) combined with
(\ref{eq85}) to read as
\begin{equation}
\label{eq86}
i_{\rm Ca} = \left[k_{\rm Ca}\,(1-x)\,d_{\infty} + k_{\rm b,Ca}
\right]\,\sinh\left(\frac{v-v_{\rm Ca}}{v_T}\right) \;,
\end{equation}

\noindent
with $v_T=kT/e$. In our model $k_{\rm b,Ca}$ is responsible for the
slow diastolic depolarization, and it is thus our pacemaker
current. However, as suggested by Guo {\rm et al.} (1995b), it is
indeed possible that the sustained inward current $i_{\rm st}$ may
largely replace the role of the ${\rm Ca}^{2+}$ currents, assumed here
and in previous studies (Wilders 1993).

\subsection{Adjustable Parameters}
The density of ionic channels, exchangers and pumps (i.e. $k_{\rm
Ca}$, $k_{\rm b,Ca}$, $k_{\rm Na}$, $k_{\rm K}$, $k_{\rm NaK}$ and
$k_{\rm NaCa}$) can vary significantly from cell to cell. In order to
reproduce recorded action potentials (Fig. 7 A. in Baruscotti {\rm et
al.}  (1996)), we fit the adjustable parameters (table \ref{table3})
and the initial conditions (table \ref{table4}) numerically. More
details of the method are given in (Endresen 1997a). Many different
combinations of $k_{\rm Ca}$, $k_{\rm b,Ca}$, $k_{\rm Na}$, $k_{\rm
K}$, $k_{\rm NaK}$, and $k_{\rm NaCa}$ resulted in good approximations
to the experimentally recorded waveform, from which we conclude that
different cells can produce the same action potential although they
have different mixtures of ionic channels, exchangers, and pumps. 

\subsection{Simulation Results}
The five differential equations in the model were solved numerically
using a fifth--order Runge--Kutta method with variable steplength.
More details are given in (Endresen 1997b). We computed the work $W$,
defined as minus the integral on the right hand side of (\ref{eq70}),
to check that the equation $W+P=0$ was satisfied numerically. This
could also be used for varying the steplength in an efficient way
(Marthinsen {\rm et al.} 1997), since the solution of our differential
equations must satisfy this constraint. These ``checksum equations''
are shown in Appendix B.

In Fig. 1 (a) the modeled action potential is shown together with the
experimental curve of Baruscotti {\rm et al.} (1996). The curves are
identical in shape, but we adjusted the modeled curve somewhat (we
multiplied the voltage amplitude by a factor 1.25, without changing
the minimum value) to obtain the same voltage amplitudes. At present
it is not clear to us which mechanism is needed in the model in order
to avoid this factor. Fig. 1 (b) shows the five membrane currents in
the model, $i_{\rm Ca}$, $i_{\rm Na}$, $i_{\rm K}$, $i_{\rm NaK}$, and
$i_{\rm NaCa}$.

Fig. 2 shows the spontaneous action potentials together with the
intracellular ionic concentrations and the osmotic pressure across the
cell membrane. These computations used the initial conditions stated
in table \ref{table4}. Cells must generate their membrane potential by
actively transporting ions against the respective concentration
gradients. To examine this process in our model, we ran a simulation
starting with equal intracellular and extracellular ionic
concentrations: ${[\rm K]_{\rm i}} = {[\rm K]_{\rm e}} = 5.4 \, {\rm
mM}$, ${[\rm Ca]_{\rm i}} = {[\rm Ca]_{\rm e}} = 2 \, {\rm mM}$, and
${[\rm Na]_{\rm i}} = {[\rm Na]_{\rm e}} = 140 \, {\rm mM}$. The
results are presented in Fig. 3, that shows  the voltage and
Nernst potentials (a), and the energies (b) in a long time simulation.
After approximately 750 seconds (12.5 minutes) the system reaches
oscillations identical to the original oscillations shown in Figs. 1
and 2 (this can not be seen from Fig. 3 since the time scale is very
different). This long time simulation is a numerical indication that
the oscillations in Fig. 2 and 3 indeed correspond to a stable limit
cycle.

\section{Discussion} 
We have presented a simple model for the cells of the rabbit
sinoatrial node. Our model involves only ${\rm Na}^{+},{\rm K}^{+}$,
and ${\rm Ca}^{2+}$ ions, their respective channels, the ${\rm
Na}^{+},{\rm Ca}^{2+}$ exchanger, and the ${\rm Na}^{+},{\rm K}^{+}$
pump. The equations were derived using basic physical principles and
conservation laws. Since the only source of energy in our model is the
sodium potassium pump, we can easily track the flow of energy, and
show that the pump works to generate a transmembrane voltage, osmotic
pressure difference, and an entropy. Our equations also account for
the energy lost due to downhill ionic fluxes through the exchanger and
channels. A prediction of osmotic pressure variations is a novel
result of our energy analysis.
 
The intracellular ionic concentrations are dynamic variables in our
model, governed by the conservation Eqs.~(\ref{eq53}), (\ref{eq54}),
and (\ref{eq55}). This allows us to replace the standard differential
equation for the voltage (\ref{eq52}) with the algebraic
Eq.~(\ref{eq63}). Although a number of other ionic models also keep
track of intracellular ionic concentrations (see Wilders (1993)), we
are unaware of any other model using an algebraic equation for the
membrane potential. Models that use the standard voltage differential
Eq.~(\ref{eq52}) have a phase space with one superfluous extra
dimension. The initial condition for this extra differential equation
cannot be chosen independently of the initial conditions for the
conservation Eqs.~(\ref{eq53}), (\ref{eq54}), and (\ref{eq55}) -- if
it is, the computed membrane potential will be erroneous. For these
reasons, we suggest that our algebraic expression for the membrane
potential should replace the standard voltage differential equation in
models where intracellular ionic concentrations are dynamic variables.
 
Our model does not include the funny current ($i_{\rm f}$), ATP
sensitive channels, stretch-activated channels, or other ion channels
that may be important (Boyett 1996). We also ignored the effect of
calcium uptake and release from the sarcoplasmatic reticulum, which
would affect the Nernst potential of calcium, but not the membrane
potential. We have assumed that the ionic channels are governed by a
Markov process, that the maximum of the activation/inactivation time
constant occurs at the same voltage as the inflection point of the
sigmoidal steady state activation/inactivation curve, and that the
steady state activation/inactivation curves were temperature
independent. Also, we have assumed that the cell volume is
constant. While such assumptions reduce the number of parameters in
the model, they may also result in discrepancies with experiment.

Finally, we would like to point out that our model is based on
experiments where some were conducted at room temperature
(22--$24^{\circ} {\rm C}$) (Baruscotti {\rm et al.} 1996; Muramatsu
{\rm et al.} 1996), while others were performed at $37^{\circ} {\rm
C}$ (Shibasaki 1987; Hagiwara {\rm et al.} 1988; Sakai {\rm et al.},
1996). It is not clear what effect varying temperature has in our
model as this was not checked out numerically.

The values of the parameters $k_{\rm Ca}$, $k_{\rm Na}$, $k_{\rm K}$,
$k_{\rm NaK}$ and $k_{\rm NaCa}$, given in table \ref{table3}, are
only an estimate of the actual physiological parameters. We did not
systematically study the dynamics of the model for different values of
the parameters, but we hope that future experiments will help to
discriminate between different parameter sets that may reproduce the
experimentally recorded action potentials.

\acknowledgements 
\noindent
{\footnotesize We are most grateful to Drs.\ Baruscotti, DiFrancesco,
and Robinson who supplied us with recordings of action potential
waveforms from rabbit sinoatrial node. Lars Petter Endresen would like
to thank professor Per Jynge for giving a fascinating introduction to
the exciting field of cardiac electrophysiology. Discussions with Per
Hemmer, K{\aa}re Olaussen, and Jacques Belair have been
essential. Lars Petter Endresen was supported by a fellowship at NTNU,
and has received support from The Research Council of Norway
(Programme for Supercomputing) through a grant of computing
time. Kevin Hall acknowledges support from the Medical Research
Council of Canada.}

\begin{description}
\item{Arrhenius SA} (1902) Text--book of Electrochemistry.
Longmans London pp 152--156.

\item{Baruscotti M, DiFrancesco D, Robinson RB} (1996) A
TTX--sensitive inward sodium current contributes to spontaneous
activity in newborn rabbit sino--atrial node cells.  Journal of
Physiology (London) 492:21--30

\item{Boyett MR, Harrison SM, Janvier NC, McMorn SO, Owen JM, Shui Z}
(1996) A list of vertebrate cardiac ionic currents: Nomenclature,
properties, function and cloned equivalents. Cardiovascular Research
32:455--481

\item{Boltzmann L} (1868) Studien {\"u}ber das Gleichgewicht der
lebendigen Kraft zwischen bewegten materiellen Punkten. Akademien der
Wissenschaften zu Berlin, G{\"o}ttingen, Leipzig, M{\"u}nchen und Wien
58:517--560

\item{Chapman JB} (1978) The reversal potential for an electrogenic
sodium pump.  A method for determining the free energy of ATP
breakdown? Journal of General Physiology 72:403--408

\item{Denyer JC, Brown HF} (1990) Calcium `window' current in rabbit
sino--atrial node cells.  Journal of Physiology (London) 429:21P

\item{DiFrancesco D} (1993) Pacemaker Mechanisms in Cardiac
Tissue. Annual Review of Physiology 55:455--472

\item{Donnan FG} (1911) Theory of membrane equilibria and membrane
potentials in the presence of non--dialysing electrolytes. A
contribution to physical--chemical physiology. Zeitschrift f{\"u}r
Elektrochemie and angewandte physikalische Chemie 17:572--581

\item{Ehrenstein G, Lecar H} (1977) Electrically gated ionic channels
in lipid bilayers. Quarterly Reviews of Biophysics 10:1--34

\item{Einstein A} (1905) {\"U}ber die von der molekularkinetischen
Theorie der W{\"a}rme gerforderte Bewegung von in ruhenden
Fl{\"u}ssigkeiten suspendierten Teilchen. Annalen Der Physik Leipzig
17:549--560

\item{Encyclop{\ae}dia Britannica 15 th edition (1997) Macrop{\ae}dia
24. p 790

\item{Endresen LP} (1997a) Chaos in weakly--coupled pacemaker cells.
Journal of Theoretical Biology 184:41--50

\item{Endresen LP} (1997b) Runge--Kutta formulas for cardiac
oscillators.  Theoretical Physics Seminars in Trondheim. No 15 ISSN
0365--2459

\item{Fick A} (1855) Ueber Diffusion. Poggendorff's Annalen der Physik
und Chemie 94:59--86

\item{Goldman DE} (1943) Potential, impedance, and rectification in
membranes. Journal of General Physiology 26:37--60

\item{Guo J, Ono K, Noma A} (1995) A sustained inward current
activated at the diastolic potential range in rabbit sinoatrial node
cells. Journal of Physiology (London) 483:1--13

\item{Guo J, Ono K, Noma A} (1995) A low--threshold sustained inward
current activated at the diastolic potent ial range in rabbit
sinoatrial node cells. Heart and Vessels 9:200--202

\item{Guo J, Ono K, Noma A} (1996) Monovalent cation conductance of
the sustained inward current in rabbit sinoatrial node
cells. Pfl{\"u}ger Archiv--European Journal of Physiology 433:209--211

\item{Hagiwara N, Irisawa H, Kameyama M} (1988) Contribution of two
types of calcium currents to the pacemaker potentials of rabbit
sino--atrial node cells. Journal of Physiology (London) 395:233--253

\item{Hille B} (1992) Ionic channels of excitable
membranes. Sunderland Massachusetts pp 127--130

\item{Hodgkin AL, Huxley AF} (1952) A quantitative description of
membrane current and its application to conduction and excitation in
nerve. Journal of Physiology (London) 117:500--544

\item{Jones HC} (1899) The modern theory of solution. Harper \&
Brothers New York and London pp 47

\item{Markov AA} (1906) Extension de la loi de grands nombres aux
{\'e}v{\'e}nements dependants les uns de autres. Bulletin de La
Soci{\'e}t{\'e} Physico--Math{\'e}matique de Kasan 15:135--156

\item{Marthinsen A, MuntheKaas H, Owren B} (1997) Simulation of
ordinary differential equations on manifolds: Some numerical
experiments and verifications. Modeling Identification and
Control}. 18:75--88

\item{Mullins LJ} (1977) A Mechanism for Na/Ca Transport. Journal of
General Physiology 70:681--695

\item{Mullins LJ} (1981) Ion Transport in Heart. Raven Press New York
pp 42

\item{Muramatsu H, Zou AR, Berkowitz GA, Nathan RD} (1996)
Characterization of a TTX--sensitive ${\rm Na}^{+}$ current in
pacemaker cells isolated from the rabbit sinoatrial node. American
Journal of Physiology 270:H2108--H2119

\item{Nernst W} (1888) Zur Kinetik der in L{\"o}sung befindlichen
K{\"o}rper. Zeitschrift f{\"u}r physikalische Chemie 3:613--637

\item{Nonner W, Eisenberg B} (1998) Ion permeation and glutamate
residues linked by Poisson--Nernst--Planck theory in L--type calcium
channels. Biophysical Journal 75:1287-1305

\item{Noma A} 1996. Ionic Mechanisms of the Cardiac Pacemaker
Potential. Japanese Heart Journal 37:673--682

\item{Ohm GS} (1827) The galvanic circuit investigated mathematically.
Berlin pp 140

\item{Onsager L} (1931) Reciprocal relations in irreversible processes
I.  Physical Review 37:405--426

\item{Sakai R, Hagiwara N, Matsuda N, Kasanuki H, Hosoda S} (1996)
Sodium--potassium pump current in rabbit sino--atrial node cells.
Journal of Physiology (London) 490:51--62

\item{Shibasaki T} (1987) Conductance and kinetics of delayed
rectifier potassium channels in nodal cells of the rabbit
heart. Journal of Physiology (London) 387:227--250

\item{van't Hoff JH} (1887) Die Rolle Des Osmotischen Druckes in der
Analogie zwischen L{\"o}sungen und Gasen.  Zeitschrift f{\"u}r
physikalische Chemie 1:481--508

\item{Varghese A, Sell GR} (1997) A conservation principle and its
effect on the formulation of Na--Ca exchanger current in cardiac
cells. Journal of Theoretical Biology 189:33--40

\item{Wilders R} (1993) From single channel kinetics to regular
beating. A model study of cardiac pacemaking activity. PhD
Thesis. pp 17--40. Universiteit van Amsterdam. ISBN 90--9006164--9

\item{Zaza A, Micheletti M, Brioschi A, Rocchetti M} (1997) Ionic
currents during sustained pacemaker activity in rabbit sino--atrial
myocytes. Journal of Physiology (London) 505:677--688

\end{description}

\newpage
\setborder[17.2cm, 1em, \smallskipamount]
\border{\appendix 
\section{Equations of Motion}
\begin{eqnarray}
v             &=& \frac{FV}{C}\left\{[\rm K]_{\rm i}-[\rm K]_{\rm e}+2([\rm Ca]_{\rm i}-[\rm Ca]_{\rm e})+[\rm Na]_{\rm i}-[\rm Na]_{\rm e}\right\} \\
v_{\rm K}     &=& v_T\ln \frac{[{\rm K}]_{\rm e}}{[{\rm K}]_{\rm e}},\;\;\; v_{\rm Ca}  = \frac{v_T}{2}\, \ln \frac{[\rm Ca]_{\rm e}}{[{\rm Ca}]_{\rm i}},\;\;\;v_{\rm Na}  = v_T\ln \frac{[\rm Na]_{\rm e}}{[{\rm Na}]_{\rm i}} \\
\; \nonumber\\
i_{\rm K}     &=& k_{\rm K}\,x\sinh\left(\frac{v-v_{\rm K}}{2v_T}\right) \\
i_{\rm Ca}    &=& \left[k_{\rm Ca}\,(1-x)\,d_{\infty} + k_{\rm b,Ca} \right]\,\sinh\left(\frac{v-v_{\rm Ca}}{v_T}\right),\;\; d_{\infty} = \frac{1}{2}\left\{1+\tanh\left(\frac{v-v_{\rm d}}{v_T/2}\right)\right\} \\
i_{\rm Na}    &=& k_{\rm Na}\,h \,m_{\infty}\,\sinh\left(\frac{v-v_{\rm Na}}{2v_T}\right),\;\; m_{\infty} = \frac{1}{2}\left\{1+\tanh\left(\frac{v-v_{\rm m}}{v_T/2}\right)\right\} \\
i_{\rm NaK}   &=& k_{\rm NaK}\,\tanh\left(\frac{v+2v_{\rm K}-3v_{\rm Na}-v_{\rm ATP}}{2v_T}\right) \\
i_{\rm NaCa}  &=& k_{\rm NaCa}\, \sinh\left(\frac {v- 3v_{\rm Na}+ 2v_{\rm Ca}}{2v_T}\right) \\
\; \nonumber\\
\frac{d}{dt}{[\rm K]_{\rm i}} &=& \frac {2i_{\rm NaK} - i_{\rm K}}{FV} \\
\frac{d}{dt}{[\rm Ca]_{\rm i}} &=& \frac {2{i_{\rm NaCa}}- i_{\rm Ca}}{2FV} \\
\frac{d}{dt}{[\rm Na]_{\rm i}} &=& \frac {-{i_{\rm Na}}- 3i_{\rm NaK}- 3i_{\rm NaCa}}{FV} \\
\frac{dx}{dt} &=&  \frac{1}{\tau_{\rm K}} \cosh\left(\frac{v-v_{\rm x}}{v_T/2}\right)\left\{\frac{1}{2}\left[1+\tanh\left(\frac{v-v_{\rm x}}{v_T/2}\right)\right]-x\right\} \\
\frac{dh}{dt} &=&\frac{1}{\tau_{\rm Na}}\,\cosh\left({\frac {v-v_{\rm h}}{v_T/2}}\right)\left\{\frac{1}{2}\left[1-\tanh\left(\frac{v-v_{\rm h}}{v_T/2}\right)\right]-h\right\} 
\end{eqnarray}
\section{Checksum Equation: $W+P=0$}
\begin{eqnarray}
\frac{dW}{dt} &=& i_{\rm K}(v-v_{\rm K}) + i_{\rm Ca}(v-v_{\rm Ca}) + i_{\rm Na}(v-v_{\rm Na}) \nonumber \\
              & & \;\;\; + i_{\rm NaCa}(v- 3v_{\rm Na}+ 2v_{\rm Ca}) + i_{\rm NaK}(v +2v_{\rm K}-3v_{\rm Na})\\
P             &=& \frac{1}{2}Cv^2 - sT - \pi V \\
s             &=& RV \left\{[\rm K]_{\rm i}\, \ln\left({\frac{[\rm K]_{\rm e}}{[\rm K]_{\rm i}}}\right)+[\rm Ca]_{\rm i}\,\ln\left({\frac{[\rm Ca]_{\rm e}}{[\rm Ca]_{\rm i}}}\right)+[\rm Na]_{\rm i}\,\ln\left({\frac{[\rm Na]_{\rm e}}{[\rm Na]_{\rm i}}}\right)\right\} \\
\pi           &=& RT \left\{[\rm K]_{\rm i}-[\rm K]_{\rm e}+[\rm Na]_{\rm i}-[\rm Na]_{\rm e}+[\rm Ca]_{\rm i}-[\rm Ca]_{\rm e}\right\} 
\end{eqnarray}}

\newpage
\vbox{
\begin{table}[h]
\caption{Fundamental Physical Constants}
\begin{tabular}{lrc} 
\multicolumn{1}{c}{Name} &
\multicolumn{1}{c}{Value} &
\multicolumn{1}{c}{Unit} \\
\hline
$k                  $ & $  1.38065812 \cdot  10^{-20}     $ & mJ/K \\
$e                  $ & $  1.60217733 \cdot  10^{-19}   $ & C \\
$F                  $ & $  96485.30929                    $ & C/mol \\
$R=kF/e             $ & $  8314.511935                    $ & J/kmol K
\end{tabular}
\label{table1}
\end{table}}

\vbox{
\begin{table}[h]
\caption{Observed Constants}
\begin{tabular}{lrc} 
\multicolumn{1}{c}{Name} &
\multicolumn{1}{c}{Value} &
\multicolumn{1}{c}{Unit} \\
\tableline
\hline
$T                        $ & $       310.15  $ & K  \\
$[\rm K]_{e}              $ & $          5.4  $ & mM \\
$[\rm Ca]_{e}             $ & $            2  $ & mM \\
$[\rm Na]_{e}             $ & $          140  $ & mM \\
$V                        $ & $           10  $ & ${\rm 10^3 \mu m^3}$ \\
$C                        $ & $           47  $ & pF \\
$v_{\rm x}                $ & $        -25.1  $ & mV \\
$v_{\rm d}                $ & $         -6.6  $ & mV \\
$v_{\rm f}                $ & $        -25.0  $ & mV \\
$v_{\rm m}                $ & $        -41.4  $ & mV \\
$v_{\rm h}                $ & $        -91.0  $ & mV \\
$v_{\rm ATP}              $ & $         -450  $ & mV \\
$\tau= {\tau}_{\rm K} = {\tau}_{\rm Ca} = {\tau}_{\rm Na} $ & $          200  $ & ms \\
$v_T = kT/e = RT/F        $ & $      26.7268  $ & mV
\end{tabular}
\label{table2}
\end{table}}

\vbox{
\begin{table}[h]
\caption{Adjustable Parameters}
\begin{tabular}{lrc} 
\multicolumn{1}{c}{Name} &
\multicolumn{1}{c}{Value} &
\multicolumn{1}{c}{Unit} \\
\tableline
\hline
$k_{\rm Ca}               $ & $   26.2     $ & pA \\
$k_{\rm b,Ca}             $ & $    0.01645 $ & pA \\
$k_{\rm Na}               $ & $  112.7     $ & pA \\
$k_{\rm K}                $ & $   32.9     $ & pA \\
$k_{\rm NaCa}             $ & $ 1400.0     $ & pA \\
$k_{\rm NaK}              $ & $   11.46    $ & pA
\end{tabular}
\label{table3}
\end{table}}

\vbox{
\begin{table}[h]
\caption{Initial Conditions}
\begin{tabular}{lrc} 
\multicolumn{1}{c}{Name} &
\multicolumn{1}{c}{Value} &
\multicolumn{1}{c}{Unit} \\
\tableline
\hline
$x_{\rm 0}                         $ & $            0.1      $ & -- \\
$f_{\rm 0} = 1 - x_{\rm 0}         $ & $            0.9      $ & -- \\
$h_{\rm 0}                         $ & $            0.008    $ & -- \\
${[\rm K]_{\rm i}}_{\rm 0}         $ & $          130.66     $ & mM \\
${[\rm Ca]_{\rm i}}_{\rm 0}        $ & $            0.0006   $ & mM \\
${[\rm Na]_{\rm i}}_{\rm 0}        $ & $           18.7362   $ & mM 
\end{tabular}
\label{table4}
\end{table}}

\newpage
\vbox{\begin{figure}[h]
\label{fig1}
\begin{center}
\epsfxsize=1.0\textwidth
\epsffile{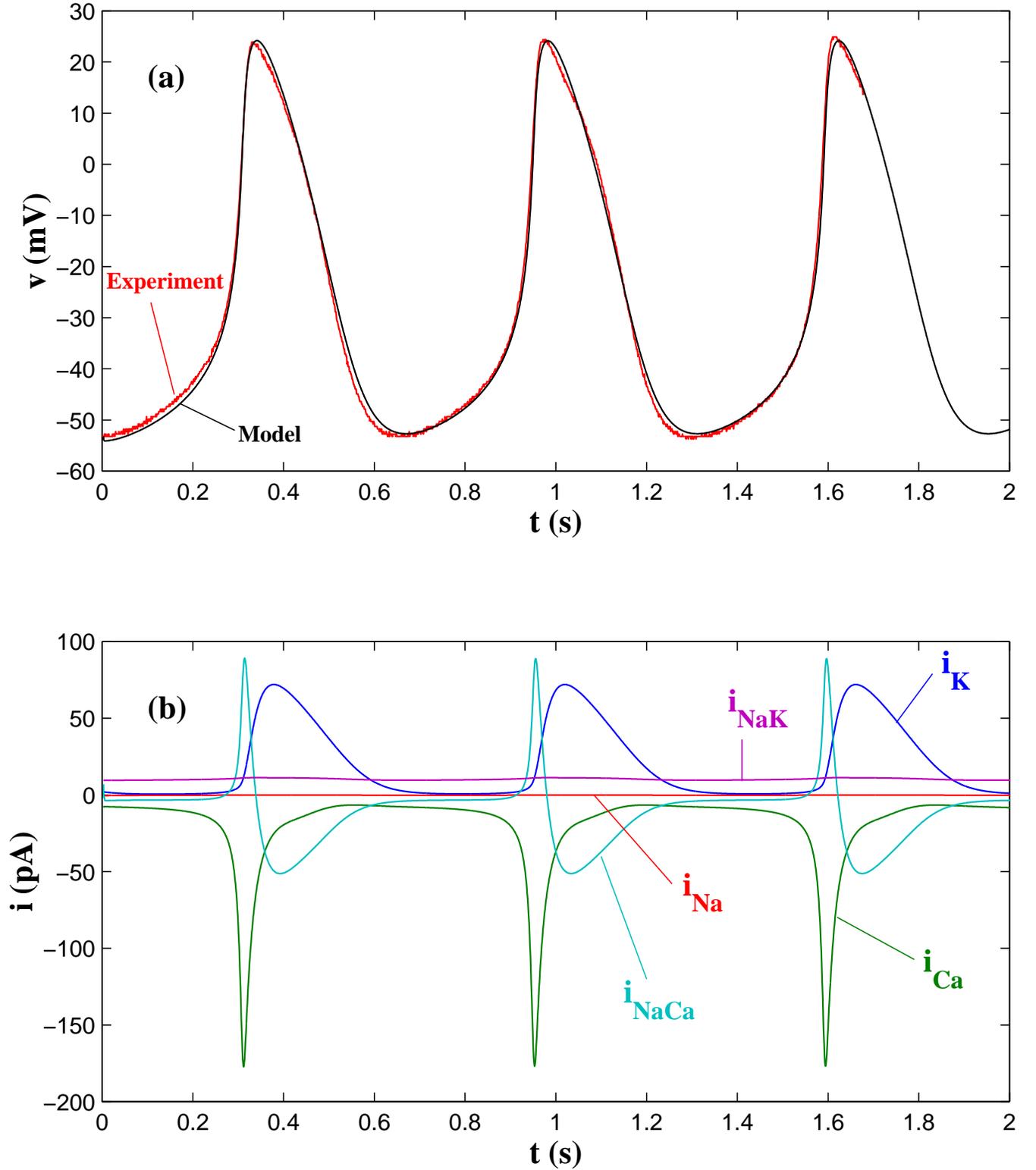}
\caption{Action potential and currents. (a) Experimentally recorded
and scaled (by a factor 1.25) model--generated rabbit sinoatrial
action potential waveform. (b) The outward delayed rectifying
potassium current ($i_{\rm K}$), the inward calcium current ($i_{\rm
Ca}$), the inward sodium current ($i_{\rm Na}$), the sodium calcium
exchange current ($i_{\rm NaCa}$) and the sodium potassium pump
current ($i_{\rm NaK}$). These computations used the initial
conditions in table \ref{table4}. }
\end{center}
\end{figure}}

\newpage
\vbox{\begin{figure}[h]
\label{fig2}
\begin{center}
\epsfxsize=1.0\textwidth
\epsffile{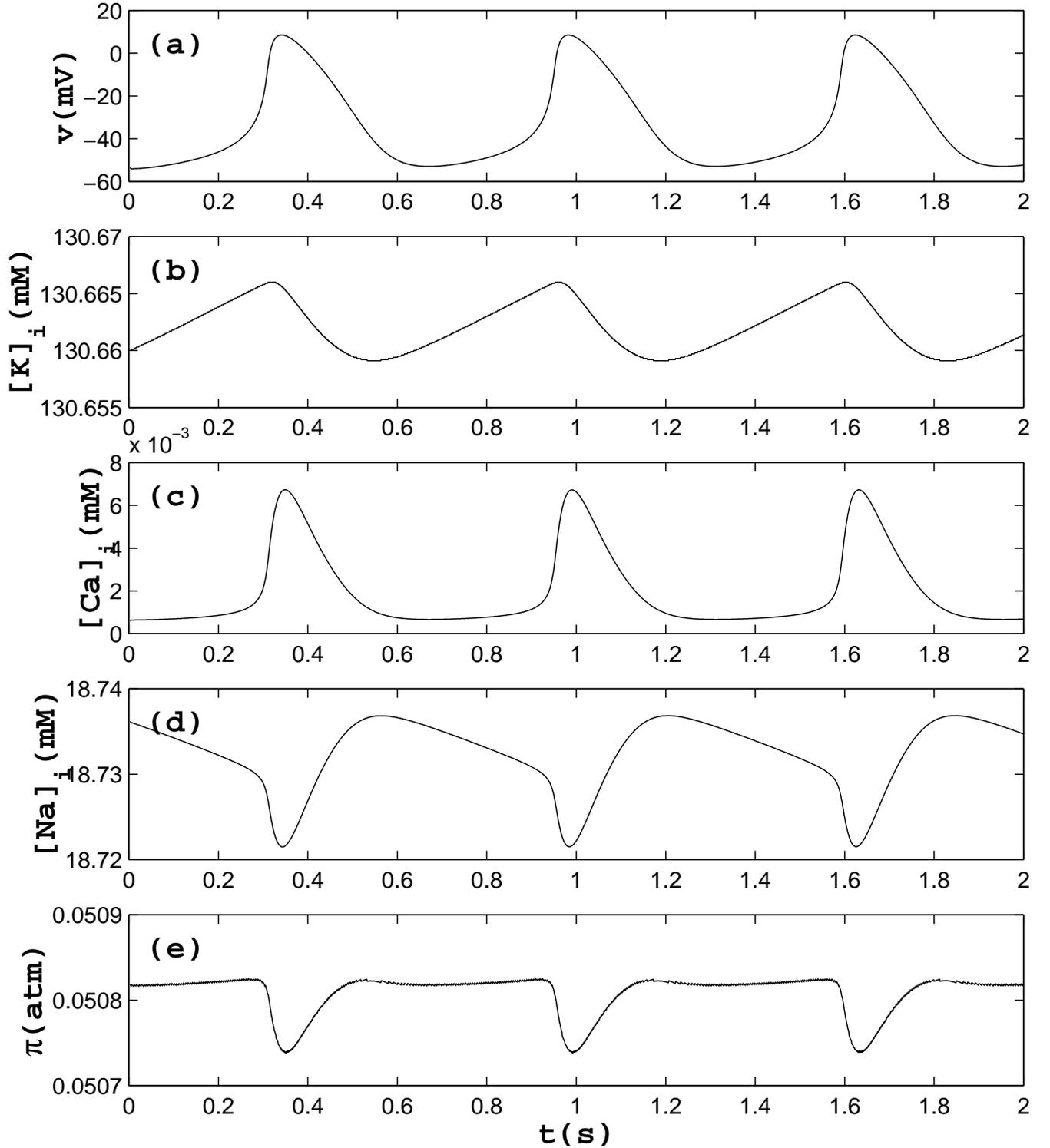}
\caption{Membrane potential (not scaled), intracellular ionic
concentrations and osmotic pressure of a rabbit sinoatrial node
cell. (a) Model--generated action potential waveform, (b) potassium
concentration $[\rm K]_{i}$, (c) calcium concentration $[\rm Ca]_{i}$,
(d) sodium concentration $[\rm Na]_{i}$ and (e) the osmotic pressure
$\pi$ across the cell membrane. These computations used the initial
conditions in table \ref{table4}.}
\end{center}
\end{figure}}

\newpage
\vbox{\begin{figure}[h]
\label{fig3}
\begin{center}
\epsfxsize=1.0\textwidth
\epsffile{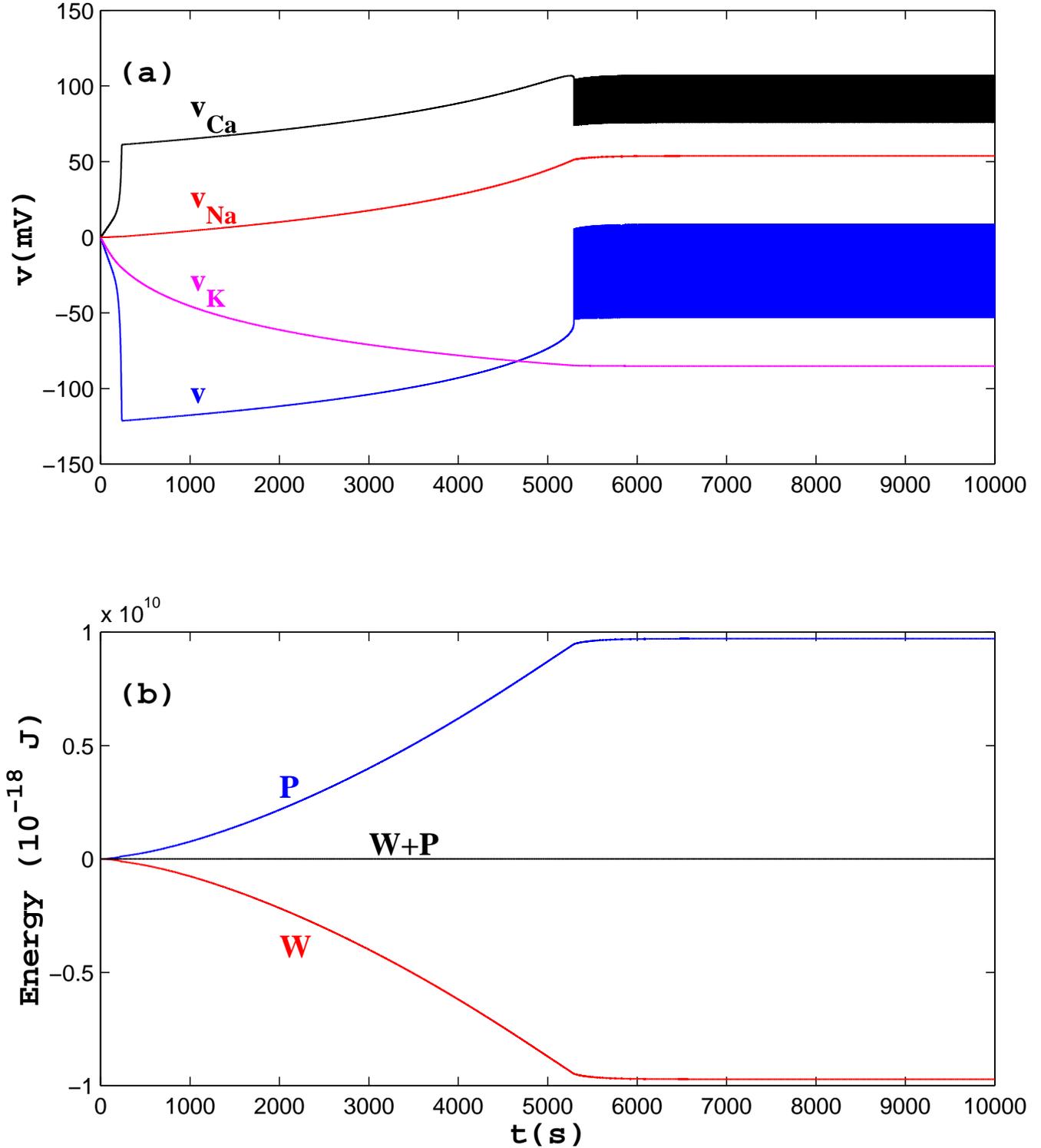}
\caption{Long time simulation showing the membrane potential, Nernst
potentials and energies starting with equal intracellular and
extracellular concentrations: ${[\rm K]_{\rm i}} = {[\rm K]_{\rm e}} =
5.4 \, {\rm mM}$, ${[\rm Ca]_{\rm i}} = {[\rm Ca]_{\rm e}} = 2 \, {\rm
mM} $ and ${[\rm Na]_{\rm i}} = {[\rm Na]_{\rm e}} = 140 \, {\rm mM}
$. (a) Nernst potential for calcium ($v_{\rm Ca}$), potassium ($v_{\rm
K}$), sodium ($v_{\rm Na}$), and membrane potential $v$.  (b) Work
($W$), potential energy ($P$) and total energy balance ($W+P$).}
\end{center}
\end{figure}}

\end{document}